\documentclass[lettersize,journal]{IEEEtran}
\usepackage{amsmath,amsfonts}
\usepackage{algorithmic}
\usepackage{algorithm}
\usepackage{array}
\usepackage[caption=false,font=normalsize,labelfont=sf,textfont=sf]{subfig}
\usepackage{textcomp}
\usepackage{bm}
\usepackage{amssymb}
\usepackage{amsthm}
\usepackage{stfloats}
\usepackage{booktabs} 
\usepackage{array}
\usepackage{verbatim}
\usepackage{tabularx}
\usepackage{multirow}
\usepackage{url}
\usepackage{verbatim}
\usepackage{graphicx}
\usepackage{cite}
\usepackage{soul}
\soulregister\cite7
\usepackage{color, xcolor}

\newtheorem{mythm}{Theorem}

\newtheorem{myass}{Assumption}
\newtheorem{myrem}{Remark}

\begin{document}

\title{Online Optimization of DNN Inference Network Utility in Collaborative Edge Computing}

\author{Rui Li$^{*}$, Tao Ouyang$^{*}$, Liekang Zeng, Guocheng Liao, Zhi Zhou, Xu Chen
\thanks{R. Li, T. Ouyang, L. Zeng, Z. Zhou, and X. Chen are with School of Computer Science and Engineering, G. Liao is with School of Software Engineering, Sun Yat-sen University, China. (e-mail: \{lirui223, zenglk3\}@mail2.sysu.edu.cn, \{ouyt33, liaogch6, zhouzhi9, chenxu35\}@mail.sysu.edu.cn). ($^{*}$These authors contributed equally to this work.) (Corresponding author: Xu Chen.)}}



\maketitle

\begin{abstract}
Collaborative Edge Computing (CEC) is an emerging paradigm that collaborates heterogeneous edge devices as a resource pool to compute DNN inference tasks in proximity such as edge video analytics.
Nevertheless, as the key knob to improve network utility in CEC, existing works mainly focus on the workload routing strategies among edge devices with the aim of minimizing the routing cost, remaining an open question for joint workload allocation and routing optimization problem from a system perspective. To this end, this paper presents a holistic, learned optimization for CEC towards maximizing the total network utility in an online manner, even though the utility functions of task input rates are unknown a priori. In particular, we characterize the CEC system in a flow model and formulate an online learning problem in a form of cross-layer optimization. We propose a nested-loop algorithm to solve workload allocation and distributed routing iteratively, using the tools of gradient sampling and online mirror descent.
To improve the convergence rate over the nested-loop version, we further devise a single-loop algorithm. Rigorous analysis is provided to show its inherent convexity, efficient convergence, as well as algorithmic optimality.
Finally, extensive numerical simulations demonstrate the superior performance of our solutions.
\end{abstract}


\section{Introduction}
Recent years have witnessed a growing explosion in the number of mobile and IoT devices \cite{Ericsson}. As foretasted by Cisco \cite{Cisco}, 29.3 billion edge devices will connect to the Internet by 2023. Meanwhile, benefited from hardware upgrades (e.g., AI accelerator chip development \cite{VenkataramaniSW21}), massive edge devices, even small embedded devices, come with deployments of emerging real-time intelligent computing tasks, such as security monitoring for timely video analytics \cite{HuangG22}. Nevertheless,  to serve these computation-intensive tasks with constrained onboard resources remains inferior performance.
Unfortunately, their inherent mission-critical nature also makes the traditional cloud offloading paradigm unsatisfactory due to expensive yet limited back-haul bandwidth and undesirably long communication distance.

\begin{figure}[t]
        \centering
	\includegraphics[scale=0.5]{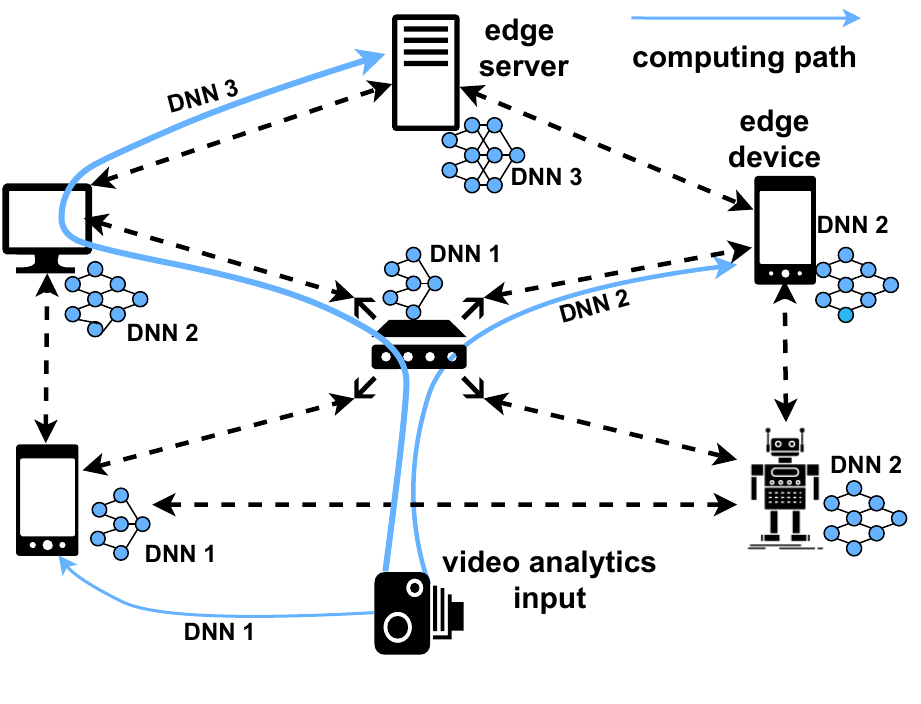}
	\caption{Illustration of CEC system topology. The edge devices are interconnected via LANs, and each device deploys a given version of DNN model to perform a specific computing task. These devices can collaborate their computation and communication resources to compute a specific application in order to achieve total network utility maximization.}
\end{figure}

To address these limitations, Collaborative Edge Computing (CEC) \cite{tran2017collaborative} has emerged as a promising solution in multi-device edge networks.
As illustrated in Fig. 1, CEC employs multiple stakeholders (e.g., IoT devices, vehicles, robots, and edge servers) to form a multi-hop network and is committed to collaborating them for shared, targeted computing tasks \cite{sahni2020multi}.
Their collaboration is at dual levels. (i) \emph{Computation collaboration}: CEC can improve the resource utilization efficiency where an edge device can utilize its neighbor’s vacant resources to complete computing tasks through
D2D communication.
Edge servers can also play as central role to orchestrate multiple edge devices for load balancing in the network level. 
(ii) \emph{Communication collaboration}: CEC can bridge multiple edge devices that cannot connect the edge server, which is indispensable in many edge scenarios. For example, rescue robot swarms can share their sensory information and computation to expedite disaster relief \cite{hong2019multi}.

As we recognize the benefits of CEC in improving resource utilization among edge devices, we should note that both the workload allocation and routing strategies play crucial roles on the CEC performance. Many existing works, however, only focus on the workload routing in CEC, where the system input rates of the multiple computing tasks are known a priori (i.e., the workload allocation is fixed advance when making routing strategies) \cite{zhang2022optimal, liu2020distributed, 9623540, sahni2018data}.
However, from a network operator's perspective, serving the computing tasks can also bring revenue related to the input rates of tasks. Therefore, it is crucial to control the task input rates to balance the cost and revenue. 

Only a few works focus on both workload allocation and routing in CEC \cite{neely2008fairness, xi2008node, jingzong2023cross}, where the revenue of the computing tasks are also taken into account. The objective of this line of work is to control the task input rates to maximize the task utility while at the same time to route workload appropriately to minimize the network cost.  However, the utility functions of task input rates are known in advance which cannot capture the various criteria of emerging mobile applications, such as inference latency or user satisfaction level. Besides, the functional relationships between the task input rates and corresponding task utility values are usually unknown for more complicated applications in practice.

Moreover, whether considering only routing or joint allocation and routing, these works do not well distinguish computing tasks with different computation and storage capacities and assume that the edge devices can process the whole kinds of computing tasks. However, this is impractical due to the intrinsically limited resources of the edge devices. Specifically, processing a computing task such as video resolution enhancement requires the edge devices to deploy corresponding DNN models, which takes up a lot of their storage space. Therefore, it is fallacious to assume every resource-limited edge device can support every computing task. Hence, the joint optimization of workload allocation and routing under unknown utility function in resource-constrained CEC environment still remains an open problem.

Inspired by the above considerations, in this paper, we consider a novel CEC framework where the multiple edge devices collaborate to finish a type of DNN inference task. 
Specially, this type of computing task is allowed to be served by a set of heterogeneous DNN models featuring diverse performance and resource requirements (different model architectures: CNN \cite{howard2017mobilenets} or Transformer \cite{DBLP:journals/corr/VaswaniSPUJGKP17}, multiple downsized version of the same pre-trained model \cite{9043731}). 
As an instance, consider a video resolution enhancement task served  by DNN models of different output-resolution versions (720P, 1080P or 2K).
Since different model versions require different storage and computing resources, and their induced utilities (e.g. quality of experience) are also different. 
However, due to limited onboard resources, each edge device can only afford a single-version DNN model deployment and need to collaborate to serve the whole computing task. 

We consider a joint optimization of workload allocation and routing to maximize the total network utility in the proposed CEC framework.
First, given a total computing task input request rate, we need to allocate the amount of workload to be processed with different DNN model versions to maximize the total task utility.
Here the workload for a specific DNN model is actually the input rate of the computing task served by the targeted DNN model.
Different from existing works in CEC, our proposed framework considers utility functions unknown a priori, which is able to capture the complex relationship between the various criteria of the DNN inference task and its allocated input rate.
After determining the task input rates, we also need to route each allocated workload among the heterogeneous edge devices optimally to minimize the total network cost produced from the communication and computation resources consumption. 
Finally, with the aim of  maximizing the total network utility, it is crucial to balance the total task utility and network cost.
Nonetheless, solving this joint optimization problem faces following tough challenges.
\begin{enumerate}
	\item \emph{Firstly, how to allocate the task input rates among heterogeneous edge devices with the unknown task utility functions?} Traditional wisdoms usually regard the utility function as a determined criterion known a priori. Nevertheless, in our considered CEC scenarios, the network utility function can be agnostic. For an emerging application, such as video analytics, the task utility usually represents the concrete quantities such as user satisfaction level \cite{hajiesmaili2012content} or inference latency. We often do not have the prior knowledge of these utility functions, i.e., the functional relationships between the task input rates and corresponding task utility values are unknown in advance.
	This makes the targeted task utility maximization quite difficult.
	
	\item \emph{Secondly, how to route each workload across
		heterogeneous edge devices under a large yet dynamic network environment?} As a key component to maximize network utility, routing computing workload within the CEC regime can effectively improve the resource utilization with respect to both computation and communication. However, the centralized routing among the massive edge devices is overwhelming, along with the topology of the CEC usually changes due to the mobility of edge devices, which urgently requires an efficient distributed routing algorithm.
		
	\item \emph{Thirdly, given the complex nexus of workload allocation and routing, how to jointly optimize them for network utility maximization?} The optimizations of workload allocation and routing are intersected with each other. Instinctively, higher admitted task input rates brings bigger task utility values, which in turn causes higher network cost since it requires more communication and computation resources to serve the admitted workload. Therefore, it is crucial to balance the task utility and network cost with the aim of maximizing the total network utility.
\end{enumerate}

To cope with the challenges, we study the joint optimization on workload allocation and routing as a max-min Network Utility Maximization (NUM).
Given its intractable hardness brought by the lack of priori knowledge, we propose a series of online learning algorithms inspired by cross-layer optmization \cite{jin2004fast} in TCP.
Specifically, our solution operates in a nested-loop manner:
in the outer loop, we apply gradient sampling to estimate the gradient of the total network utility, and then propose an online mirror ascent technique to iterate the allocation decisions;
in the inner loop, we apply the online mirror descent algorithm again to solve an optimal routing under previously determined allocated task input rates.
In this way, we can distributively iterate workload allocation and workload routing until a convergence is reached.
To expedite its converge rate, we further devise an improved single-loop algorithm upon the nested-loop one, based on its mathematical performance analysis.
Rigorous proofs as well as numerical simulations show the problems' convexity, as well as our solution's efficient convergence and algorithmic optimality.

Our key contributions can be summarized as follows:
\begin{enumerate}
	\item We study a novel Network Utility Maximization (NUM) in Collaborative Edge Computing for DNN inference tasks scenarios that allows the task utility functions unknown a priori.
	We establish a novel Joint Optimization framework on Workload allocation and Routing (JOWR) problem formulation.
	\item We propose a cross-layer online optimization framework to tackle the JOWR problem. Specifically, we design a nested-loop algorithm to solve workload allocation and workload routing in two time scales, and iterate them for a provably converged result. Tailored to the decision space geometry, our algorithm exhibits a fast convergence speed, while requiring less communication and computation overhead.	
	\item Based on the rigorous performance analysis, we further devise a single-loop algorithm to improve the convergence rate upon the nested-loop counterpart, with a provable efficient convergence guarantee.
	\item We conduct extensive numerical simulations in various settings, demonstrating the superior performance of the proposed solution with respect to both the nested- and single-loop algorithms.
\end{enumerate}

The rest of this paper is organized as follows. We first present the system model and problem formulation in Section~\ref{problem}. Then we propose a nested-loop algorithms to solve the JOWR problem at two timescales and theoretically analyze the online performance in Section~\ref{online optimization}. Next, we carry out performance evaluation in Section~\ref{performance eval}, and finally conclude in Section~\ref{conc}.
\section{Problem Formulation}
\label{problem}
In this section, we first introduce the collaborative edge computing model where multiple edge devices collaborate to finish a DNN inference task. Note that our system model can be adapted to many real-world applications, such as robot or UAV swarms, smart home or city, and IoT networks. Through heterogeneous resource sharing, these devices can collaborate to compute some specific intelligent applications in order to achieve total network utility maximization. For ease of exposition, the main notations used in this paper and their physical meanings are summarized in Table 1.

\subsection{Network and Computation Models}
We consider a multihop wireless network, denoted by a directed and (strongly) connected graph $\mathcal{G}=(\mathcal{N}, \mathcal{E})$, where $\mathcal{N}$ and $\mathcal{E}$ are the sets of nodes and links, respectively. A node $i \in \mathcal{N}$ corresponds to an edge device that contains some computation and communication resources. A link $(i, j) \in \mathcal{E}$ represents a directed link, in which node $i$ can transmit the computation task to node $j$. Throughout this paper, we assume that the capacity of a link $(i,j) \in \mathcal{E}$ is fixed and denote it as $C_{ij}$ (in bits/sec). In practice, when the network transport protocol and the nodes' transmission powers are determined, we can simply treat the link capacity as fixed. In practical scenarios with time-variant link capacity and random noise, our online optimization approach can still work, assuming the link capacity has a constant mean $C_{ij}$ with a zero mean noise.

To leverage the resource heterogeneity of edge devices, we allow different DNN models with different resource requirements run on the edge devices. Specifically, we consider a specific DNN inference task (e.g., face recognition or object detection) can be served with different quality levels (e.g., inference accuracy and delay) by using different versions (characterized by storage and computation requirements) of DNN models. For example, the face recognition task can be served using a series of ResNet versions with different model sizes and therefore leads to different accuracy \cite{ResNet}. Intuitively, a DNN model with bigger storage and computation requirement can achieve higher accuracy but longer latency. Let $\mathcal{W}$ be the set of DNN versions of a specific DNN inference task with cardinality $W=|\mathcal{W}|$. For simplicity, we assume that each edge device runs with one suitable DNN model version in accordance with its storage and computation resources. Particularly, in scenarios where an edge device has sufficient resources to deploy multiple DNN models, we can treat the actual device as multiple virtual devices, with each virtual device deploying only one DNN model, as illustrated in Fig. 2. This approach enables us to accommodate scenarios with multiple DNN models per node. Note that the DNN model placement is fixed and the deployed DNN model on each edge device is determined by its device user beforehand.

\begin{table}[t]
	\caption{MAIN NOTATIONS AND THEIR MEANINGS}\label{table one}
	\begin{tabular}{ll}
		\toprule
		notations	&  meanings\\
		\midrule
		$\mathcal{G}$	& The CEC network topology\\
		$\mathcal{N}$	& The set of edge devices\\
		$\mathcal{E}$	& The set of wireless links \\
		$C_{ij}$        & The capacity wireless link $(i,j)$\\
		$\mathcal{W}$   & The set of DNN versions of a specific task\\
		$\lambda$       & The DNN inference task input rate\\ 
		$\Lambda$       & The workload allocation decision\\
		$u_w$           & The unknown utility function of DNN model $w$\\
		$D_w$           & The virtual node of destination of session $w$\\
		$D(w)$          & The set of devices deploys $w$-th version of DNN model\\
		$f_{ij}(w)$     & The flow rate of traffic session $w$ on link $(i,j)$\\
		$t_i(w)$        & The total incoming request rate of session $w$ at node $i$\\
		$\mathcal{I}(i)$ & The in-coming neighbors of node $i$\\
		$\mathcal{O}(i)$ & The out-going neighbors of node $i$\\
		$\phi_{ij}(w)$  & The fraction of session $w$'s traffic at link $(i,j)$\\
		$F_{ij}$        & The total flow rate on a link $(i,j)$\\
		$D_{ij}$        & The cost function on link $(i,j)$\\
		\bottomrule
	\end{tabular}
\end{table}
\subsection{Inference Task Utility Model}
Given a DNN inference task input rate $\lambda$,\footnote{For example, the rate of generating image frames is usually fixed for a camera or a set of nearby cameras for video analytics applications.} we consider the total inference task utility as $\sum_{w \in \mathcal{W}}u_w(\lambda_w)$, where the workload allocation decision (i.e., the input rate of each task served by each DNN model) is $\Lambda \triangleq [\lambda_1,\ldots,\lambda_W]$ with $\lambda_w \ge 0$, for $w=1,2..,W$ and $\sum_{w=1}^{W}\lambda_w = \lambda$. Here, we consider a utility function $u_w(\lambda_w)$ associated with  the $w$-th DNN model, which depends on the workload $\lambda_w$ of the model. The value of the utility can be instantiated as model accuracy or inference delay. However, the relationship between the utility and the workload is unknown. We can only observe and obtain the utility $u_w(\lambda_w)$ when there is $\lambda_w$ tasks served by $w$-th DNN model. Assuming that the utility value $u_w(\lambda_w)$ represents the inference delay, we cannot predict the actual delay $u_w(\lambda_w)$ due to the unknown network condition and resources situation of the serving edge devices.
For each DNN model version, we assume that its utility function has the following properties:
\begin{myass}
$u_w$ is an monotonically increasing, continuously differentiable and concave function.
\end{myass}
\begin{myass}
$u_w$ is $L_w$-Lipschitz continuous, i.e., $\forall \lambda_w^1, \lambda_w^2 \in [0, \lambda]$, we have $|u_w(\lambda_w^2) - u_w(\lambda_w^1)| \le L_w \cdot |\lambda_w^2 - \lambda_w^1|$.
\end{myass}
 \begin{myass}
$u_w$ is bounded on $[0, \lambda]$, i.e., $\forall \lambda_w \in [0, \lambda], u_w(\lambda_w) \le B$ for some constant $B$.
 \end{myass}

These assumptions are quite common and widely adopted in existing works \cite{fu2021learning}. Moreover, the ``monotonically increasing'' assumption illustrates that the utility value strictly increases with the task input rate, which effectively captures the service performance in practical deployments, such as inference throughput \cite{WuWPCLY22}.

\begin{figure}[t]
        \centering
	\includegraphics[scale=0.5]{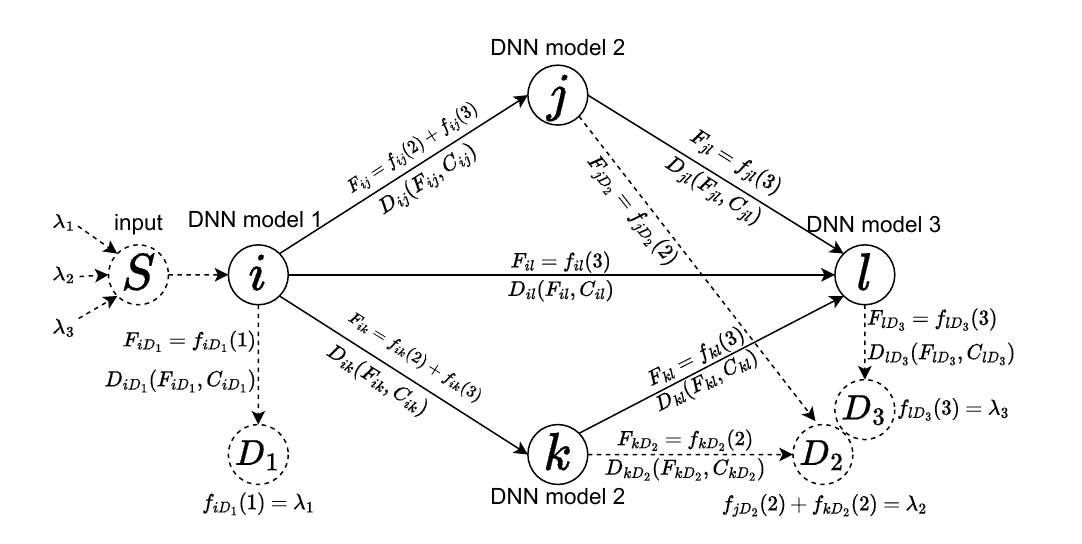}
	\caption{Session 1,2,3 with rates $\lambda_1, \lambda_2, \lambda_3$ all originate at the virtual source node $S$ with destinations to virtual nodes $D_1, D_2, D_3$, respectively. Node $i$ routes session 2 to $D_2$ through $j$ and $k$, and routes session 3 to $D_3$ through $j,k$ and $l$. The total flow rate on link $(i,j)$ is the sum of flow rate of all sessions passing through it. Finally, the total incoming rate of each virtual node $D_w$ must be equal to the session rate $\lambda_w$.}
\end{figure}
\subsection{Traffic Model}
Given the workload allocation decision $\Lambda \triangleq [\lambda_1,\ldots,\lambda_W]$, i.e., the task input rate of each version of model, we adopt a flow model \cite{bertsekas2021data} to analyze the routing of the collection $\mathcal{W}$ of traffic sessions (each session corresponding to each version of DNN model). Throughout this paper, we assume that each traffic session can be divided into arbitrarily fine partitions and forwarded to its destination in a hop-by-hop multi-path routing.

In a flow model, each traffic session $w$ is identified by its source-destination node pair. Each session might have multiple destinations, that is, there are multiple edge devices deploying the same $w$-th DNN model, and they collaborate to serve session $w$. Therefore, we add a virtual node $D_w$ to the original graph $\mathcal{G}$ as the destination of session $w$ that connects to $D(w)$, where $D(w)$ denotes the set of all edge devices which deploying the $w$-th version of DNN model. Moreover, we add a virtual source node $S$ to $\mathcal{G}$ as the common origin of all sessions and other corresponding virtual links. The augmented graph $\bar{\mathcal{G}}$ is illustrated in Fig. 2. Let $\mathcal{N}_v$ and $\mathcal{E}_v$ be the set of virtual nodes and links, respectively, where $\mathcal{N}_v=\{S, D_1, \ldots, D_W\}$, and $\mathcal{E}_v=\{(S,i):i \in D(1)\} \cup \{(i,D_w):i \in D(w), D_w \in \mathcal{N}_v \setminus S\}$. The common origin node $S$ can be a central controller to admit the total workload $\lambda$. For simplicity, we assume that the controller has the direct connection with a few edge devices (e.g., deploying the smallest DNN model version) in proximity. The DNN inference tasks can also be processed by other edge devices/nodes (e.g., deploying other larger model versions) via one-hop or multi-hop relays via some edge devices or intermediate edge nodes. For example, the task from virtual node $S$ to node $l$ must be relayed by nodes $i,j,k$ in Fig. 2. When the network is not strongly connected, we can partition the whole network topology graph into multiple clusters, and we can apply our algorithms within each cluster wherein the subnetwork is strongly connected. We denote the augmented graph as $\bar{\mathcal{G}}=(\bar{\mathcal{N}}, \bar{\mathcal{E}})$, where $\bar{\mathcal{N}}=\mathcal{N} \cup \mathcal{N}_v$, and $\bar{\mathcal{E}}=\mathcal{E} \cup \mathcal{E}_v$.

Let $f_{ij}(w)$ denote the flow rate of traffic session $w$ on link $(i,j)$. We should guarantee the following flow conservation relations in the augmented graph $\bar{\mathcal{G}}$. For all sessions $w \in \mathcal{W}$,
\begin{equation}
	\begin{aligned}
		& f_{ij}(w) \ge 0, \quad \forall (i,j) \in \bar{\mathcal{E}},\\
		\sum_{j \in \mathcal{O}(i)} & f_{ij}(w) = \lambda_w \triangleq t_i(w), \quad i = S,\\
		& f_{ij}(w) = t_i(w), \quad \forall i \in D(w)\quad \mathrm{and} \quad j = D_w,\\
		\sum_{j \in \mathcal{O}(i)} & f_{ij}(w) = \sum_{j \in \mathcal{I}(i)} f_{ji}(w) \triangleq t_i(w), \quad \mathrm{otherwise},\label{a}
	\end{aligned}
\end{equation}
where $\mathcal{I}(i) \triangleq \{j:(j,i) \in \bar{\mathcal{E}}\}$ is the in-coming neighbors of node $i$ and $\mathcal{O}(i) \triangleq \{j:(i,j) \in \bar{\mathcal{E}}\}$ in the out-going neighbors of node $i$. Here, $t_i(w)$ is defined as the session $w$'s total incoming request rate at node $i$. In this paper, we assume that a node cannot relay a task to another node with the same model, as it would introduce additional transmission costs.

Although there have been extensive studies of multi-path source routing techniques to obtain the optimal routing in a flow model \cite{wang2003optimal} \cite{lin2003multi}, they both have the assumption that all the paths to the destinations are known a prior at the source node. However, such assumptions are not suitable for the popular wireless collaborative computing scenario whose topology changes frequently as some nodes are moving. Therefore, in this paper, we focus on distributed node-based routing where all nodes perform routing only with their immediate neighbors, and are not required to know the topology of the entire network. 

We adopt the routing variables proposed by Gallager \cite{1093711} to implement the distributed scheme. Let $\phi_{ij}(w) \in [0,1]$ denotes the fraction of session $w$'s traffic forwarded to node $j$ from node $i$, that is
\begin{equation}
	\phi_{ij}(w) \triangleq \frac{f_{ij}(w)}{t_i(w)}, \quad \forall j \in \mathcal{O}(i).\label{b}
\end{equation}
Therefore, the flow conservation relations in \eqref{a} are transformed into:
\begin{equation}
	\begin{aligned}
		&\phi_{ij}(w) \ge 0, \quad \forall j \in \mathcal{O}(i)\\
		\sum_{j \in \mathcal{O}(i)}&\phi_{ij}(w) = 1, \quad \mathrm{if} \quad i \neq D_w,\\
		&\phi_{ij}(w) = 1, \quad \forall i \in D(w)\quad \mathrm{and} \quad j = D_w.\label{c}
 	\end{aligned}
\end{equation}
Let $\bm{\phi}$ denote the routing configuration and $\mathcal{H}(\bm{\phi})$ denote the decision space of routing variables. 
In the case where a node $i$ might have that $t_i(w) = 0$, the concrete values of $\phi_{ij}(w)$ are insignificant to the actual flow rates, we simply assume that $\phi_{ij}(w) = 0$ when $t_i(w)=0$ and such $\phi_{ij}(w)$ can violate the constraint \eqref{c}.

\subsection{Network Cost Model} 
We capture the two kinds of cost in the network: communication cost and computation cost. 

\textbf{Communication cost:} We denote the communication cost on a link $(i,j)$ as $D_{ij}(F_{ij}, C_{ij})$, which depends on the sum of flow rates of all the traffic sessions on the link $F_{ij}$:
\begin{equation}
	F_{ij} = \sum_{w \in \mathcal{W}}t_i(w)\phi_{ij}(w), \quad \forall (i,j) \in \bar{\mathcal{E}},\label{d}
\end{equation}
and the capacity of the link $C_{ij}$. Here, the function $D_{ij}(F_{ij}, C_{ij})$ is an increasing, continuously differentiable and convex function in the sum of flow rates $F_{ij}$ for any fixed $C_{ij}$. 
Such convex cost can represent a variety of cost metrics in practical network, e.g., linear cost \cite{ioannidis2017jointly} for energy consumption, or convex cost for queueing delay that reflects the network congestion status. For instance, 
\begin{equation}
	D_{ij}(F_{ij}, C_{ij}) = \frac{F_{ij}}{C_{ij}-F_{ij}},\quad \mathrm{for} \quad 0 \le F_{ij} < C_{ij}\label{e}
\end{equation} 
gives the expected delay at link $(i,j)$ under an $M/M/1$ queueing model. Moreover, one could relax the hard capacity constraint $F_{ij} < C_{ij}$ by any appropriate convex functions such as $D_{ij}(F_{ij},C_{ij})=\exp(a_{ij} \cdot (F_{ij}/C_{ij}))$, where $a_{ij}$ is a cost coefficient of link $(i,j)$.

\textbf{Computation cost:} We denote the computation cost on a node $i$ as $D_i(t_i(w),C_i)$, which depends on the session $w$'s total incoming request rate $t_i(w)$, and the computing capacity of node $C_i$. Note that the edge device deploying the $w$-th DNN model can only serve session $w$. Similar to the communication cost function, we assume that $D_i(\cdot, C_i)$ is also an increasing, continuously differentiable and convex function for any fixed $C_i$. Such cost function can be instantiated as the inference accuracy loss or delay for node $i$.

In the augmented graph $\bar{\mathcal{G}}$, we can simply treat the computation cost at node $i$ as the communication cost at the virtual link $(i, D_w)$, where we have the following transformations:
\begin{equation}
	D_{iD_w} = D_i, C_{iD_w} = C_i, \mathrm{\text{ and }} F_{iD_w} = t_i(w).\label{f}
\end{equation} That is, the computation cost $D_i$ turns to the communication cost $D_{iD_w}$, the computation capacity $C_i$ turns to the communication capacity $C_{iD_w}$ of virtual link $(i, D_w)$, and the served workload $t_i(w)$ turns to the flow rate $F_{iD_w}$. The network model and cost functions are illustrated in Fig. 2. Therefore, the total network cost (consisting of communication and computation cost) can be formulated as the sum of costs of all links in augmented graph $\bar{\mathcal{G}}$.

\subsection{Optimization Problem}
Now we can formulate the main Jointly Optimal Workload allocation and Routing (JOWR) problem. By adjusting
workload allocation $\Lambda$ to maximize the task utility and routing variable $\bm{\phi}$ to minimize both computation and communication cost, we aim to maximize the total network utility as follows:
\begin{align}
	\max \quad & U(\Lambda,\bm{\phi}) \triangleq \sum_{w \in \mathcal{W}}u_w(\lambda_w) - \sum_{(i,j) \in \bar{\mathcal{E}}}D_{ij}(F_{ij},C_{ij}) \label{g}\\
	\mathrm{s.t.} \quad & \forall \lambda_w \ge 0 \mathrm{\text{ and }}\sum_{w \in \mathcal{W}} \lambda_w = \lambda, \label{h}\\
	&\mathrm{\text{flow conservation constraints }} \eqref{a}-\eqref{d} \notag.
\end{align}
The total network utility equals to the inference task utility minus the network cost. Let $\mathcal{H}(\Lambda)$ denote the decision space of $\Lambda$ satisfying the constraint \eqref{h}. The most fundamental challenge of solving JOWR problem is that the two sets of decisions are coupled. We note that the above JOWR problem is kind similar to the classic works of cross-layer optimization in wireless network \cite{chiang2007layering} \cite{lin2006tutorial}, where $\Lambda$ are the congestion control variables at the transport layer and $\bm{\phi}$ are routing variables at the network layer. However, the utility function is assumed to be known in traditional cross-layer optimization works, while we consider the unknown utility function $u_w$ which is more suitable in practice. Moreover, previous works \cite{la2002utility,chiang2007layering,lin2006tutorial} mainly assume that the routing variables have linear costs with link capacity while we assume that the link costs are convex. Finally, we focus on the distributed node-based routing problem while most of them concentrated on centralized path-based routing.

\section{Cross-layer Online Optimization algorithms}
\label{online optimization}
In this section, we first propose the nested-loop algorithm to solve the JOWR problem at two timescales. In the outer loop, we iterate the workload allocation decision $\Lambda$ to accumulate the knowledge of the unknown utility function $u_w$, and finally obtain the optimal solution $\Lambda^{*}$. In the inner loop, given a fixed $\Lambda$, we iterate the routing variables $\bm{\phi}$ and obtain the optimal solution $\bm{\phi}^{*}(\Lambda)$. We present the outer loop and inner loop in Section 3.1 and Section 3.2, respectively. Then, in Section 3.3, we present a single-loop algorithm where we iterate both $\Lambda$ and $\bm{\phi}$ simultaneously to improve the convergence rate.

Based on the above description, the values of two key components in the objective function, namely the task utility $u(\cdot)$ and the network cost $D(\cdot,\cdot)$, depend respectively on the allocation $\Lambda$ and routing $\bm{\phi}$ variables. An intuitive idea is to solve the network utility optimization $\max U(\Lambda,\bm{\phi})$ by separately maximizing task utility and minimizing network cost. In the following, we further discuss the detailed implementation based on the rigorous theoretical analysis.
\subsection{Optimal Workload Allocation Algorithm}
In order to obtain the optimal workload allocation decision $\Lambda^{*}$, we need to tackle two main challenges. Firstly, the workload allocation has a critical influence on the network cost. Specifically, the workload allocation affects the optimal routing decision, and further affects the network cost. Secondly, the utility functions $u_w$ are unknown, and we can only learn their values in an online manner over the process of decision making. Further, it is non-trivial to get the hidden relationship between $\Lambda$ and $\sum_{(i,j) \in \bar{\mathcal{E}}}D_{ij}$. 

To deal with the first challenge, we need the following assumption:
\begin{myass}
	Given any allocation decision $\Lambda$, there exists an oracle algorithm $\mathfrak{O}$ that derives the optimal routing decision $\bm{\phi}^{*}(\Lambda) \in \mathcal{H}(\bm{\phi})$ which minimizes the total network cost $\sum_{(i,j) \in \bar{\mathcal{E}}}D_{ij}$.\footnote{We will show that Assumption 1 always holds when we consider the optimal routing algorithm, and derive a specific instantiation of the oracle Algorithm in Section \ref{footnote2}.}
\end{myass}
Applying assumption 4 to the JOWR problem, we now only need to solve the following optimal workload allocation problem $\mathcal{P}1$:
\begin{align}
	\mathcal{P}1: \max_{\Lambda}  U(\Lambda, \bm{\phi}^{*}(\Lambda)) &\triangleq  \sum_{w \in \mathcal{W}}u_w(\lambda_w) - \sum_{(i,j) \in \bar{\mathcal{E}}}D_{ij}(F_{ij}^{*}(\Lambda),C_{ij}) \label{i}\\
	& \mathrm{s.t.} \quad \eqref{h}, \notag
\end{align}	
where $F_{ij}^{*}(\Lambda)$ is the hidden optimal total flow rate on link $(i,j)$ when the workload allocation is $\Lambda$.

To deal with unknown utility functions $u_w$ and hidden relationship $F_{ij}^{*}(\Lambda)$, we combine the ideas of Gradient Sampling \cite{flaxman2004online} and Online Mirror Ascent \cite{shalev2012online} to propose an online workload allocation algorithm called GS-OMA that works for problem $\mathcal{P}1$. Before interpreting the details of GS-OMA, we first give the following theorem based on which we can guarantee the optimality of proposed GS-OMA and check whether the convergence is achieved by verifying the conditions stated in the theorem. The proof of Theorem 1 is deferred to Appendix A. 

\begin{mythm}
	If Assumption 4 holds, 
	problem $\mathcal{P}1$ in \eqref{i} has a unique solution $\Lambda^{*}$. The necessary and sufficient condition of optimality is $\frac{\partial U}{\partial \lambda_1^{*}} = \cdots = \frac{\partial U}{\partial \lambda_w^{*}} = \cdots = \frac{\partial U}{\partial \lambda_W^{*}} = \alpha^{*}$, where $\alpha^{*}$ is the optimal Lagrangian multiplier of the constraint $\sum_{w \in \mathcal{W}} \lambda_w = \lambda$.
\end{mythm}

\begin{myrem}
	The necessary and sufficient condition of optimality in Theorem 1 can be used as the stopping criterion of the learning process for solving problem \eqref{i}. Intuitively, the partial derivatives of each element of the optimal point are the same means that it has no incentive to change the decision at that point. 
\end{myrem}

\begin{algorithm}[thbp] 
	\caption{The optimal workload allocation algorithm GS-OMA} 
	\label{alg1} 
	\begin{algorithmic}[1] 
		\REQUIRE  
		The oracle algorithm $\mathfrak{O}$, the disturbance $\delta$, the step sizes $\eta_t$
		\ENSURE 
		The optimal workload allocation decision $\Lambda^{*}$
		\STATE \textbf{Initialize:} $\Lambda^1=\frac{\lambda}{W} \cdot \mathbf{1}$, where $\mathbf{1}$ is the all one vector with coordinate $W$
		\STATE \textbf{for} each outer loop $t = 1,\ldots, T$ \textbf{do}
		\STATE \quad \textbf{for} each session $w = 1,\ldots,W$ \textbf{do} 
		\STATE \quad \quad allocate $\Lambda^{+}(t) = \Lambda^{t} + \delta \bm{e}_w$ and invoke oracle $\mathfrak{O}$ to 
		
		\quad \quad observe $U^{+} = U(\Lambda^{+}(t),\bm{\phi}^{*}(\Lambda^{+}(t)))$
		\STATE \quad \quad allocate $\Lambda^{-}(t) = \Lambda^{t} - \delta \bm{e}_w$ and invoke oracle $\mathfrak{O}$ to 
		
		\quad \quad observe $U^{-} = U(\Lambda^{-}(t),\bm{\phi}^{*}(\Lambda^{-}(t)))$
		\STATE \quad \quad gradient estimate: $\frac{\partial U}{\partial \lambda_w^{t}}=\frac{U^{+}-U^{-}}{2\delta}$
		\STATE \quad \textbf{end for}
		\STATE \quad workload allocation update:
		
		\quad \quad \quad 
		\begin{equation}
			\lambda_w^{t+1} = \frac{\lambda_w^{t}\exp(\eta_t\frac{\partial U}{\partial \lambda_w^{t}})}{\sum_{w=1}^{W}\lambda_w^{t}\exp(\eta_t\frac{\partial U}{\partial \lambda_w^{t}})}, \quad \forall w \in \mathcal{W}	\label{s}	
		\end{equation}
		\STATE \quad projection step: $\Lambda^{t+1} = \mathcal{P}_{[\delta, \lambda-\delta]^{W}}[\Lambda^{t+1}]$
		\STATE \quad \textbf{if} $\Lambda^{t+1}==\Lambda^{t}$, \textbf{break}
		\STATE \textbf{end for}
	\end{algorithmic}
\end{algorithm}
Now, we are ready to interpret our proposed GS-OMA algorithm for optimal workload allocation problem \eqref{i}. Note that $\bm{e}_w$ is an unit vector with the $w$-th element equals to one while the others are zeros. Since the utility functions $u_w$ and hidden relationship $F_{ij}^{*}(\Lambda)$ are unknown, GS-OMA constructs the gradient estimates using observations of function values. Specifically, at each outer loop $t \in \{1,\ldots,T\}$, each session $w$ allocates a first request rate $\lambda_w^{t} + \delta$ and a second request rate $\lambda_w^{t} - \delta$ and obtains the feedback of value $U^{+}$ and $U^{-}$ (Lines 4, 5). These two feedback values are combined to form the partial gradient estimate of $U$ at $\lambda_w^{t}$ (Line 6). The estimated gradient is then fed into the update of workload allocation variables where we apply the online mirror ascent technique to maximize $U$ (Line 8). One benefit of applying the mirror descent \eqref{s} is that after gradient updating, the new decision variables $\Lambda^{t+1}$ is still in $\mathcal{H}(\Lambda)$.  The projection step $\mathcal{P}_{[\delta, \lambda-\delta]^{W}}$ of Line 9, defined as the projection onto space $[\delta, \lambda-\delta]^{W}$ by the Euclidean norm, is to ensure that each $\lambda_w^{t} + \delta$ and $\lambda_w^{t} - \delta$ always lie in the domain $[0, \lambda]$. Finally, we can stop the outer loop when the allocation $\Lambda^{t}$ does not change (Line 10).
\begin{myrem}
	 We adopt the online mirror ascent in GS-OMA to adjust gradient updates to fit the geometry of the decision space $\mathcal{H}(\Lambda)$, which would have a significance improvement on the convergence rate than the canonical online gradient ascent algorithm in practice. 
\end{myrem}
We should point out that the gradient estimation $\bm{g}^t \triangleq [\frac{\partial U}{\partial \lambda_1^{t}} \\ \cdots  \frac{\partial U}{\partial \lambda_w^{t}}]$ in Algorithm 1 is an approximated gradient of $U$ at point $\Lambda^{t}$ since the exact gradient $\nabla U(\Lambda^{t})$ requires $\delta \rightarrow 0$. However, involving the gradient error would complicate the convergence analysis dramatically. In order to concentrate on the key performance of the GS-OMA algorithm, we have the following assumption: 
\begin{myass}
	When the disturbance parameter $\delta$ is small enough, the gradient estimation $\bm{g}^t$ in Algorithm 1 is a subgradient of $U$ at point $\Lambda^{t}$. That is, $\bm{g}^t \in \partial U(\Lambda^{t})$ satisfies:
	\begin{equation}
		U(\Lambda) \le U(\Lambda^{t}) + \langle \bm{g}^t, \Lambda - \Lambda^{t}\rangle
	\end{equation}
	where $U(\Lambda)$ is the simplified notation of $U(\Lambda, \bm{\phi}^{*}(\Lambda))$.
\end{myass} 
\begin{myass} 
$U(\Lambda)$ has Lipschitz continuous gradient. Specifically, there exists an constant $L_U > 0$ such that 
\begin{equation}
	||\nabla U(\Lambda) - \nabla U(\Lambda^{t})||_{*} \le L_U ||\Lambda - \Lambda^{t}||^2, \quad \forall \Lambda, \Lambda^{t}.
\end{equation}
\end{myass}

Such smoothness of $U$ is easy to satisfy when we have well-behaved\footnote{We claim that a function is well-behaved when it is continuously differentiable and smooth.} utility function $u_w$ and cost function $D_{ij}$. Now, we can give the convergence performance of GS-OMA algorithm. 

\begin{mythm}
	Under Assumptions 4-6, the optimal workload allocation algorithm GS-OMA converges with 
	\begin{equation}
		\min_{t \in \{2,\ldots,T+1\}} \epsilon_t \le \frac{L_UR_U^2}{T},
	\end{equation}
	where $\epsilon_t \triangleq U(\Lambda^{*}) - U(\Lambda^{t})$ is the optimization error in outer loop $t$, and $R_U$ is the diameter of the decision space $\mathcal{H}(\Lambda)$.
\end{mythm}
The proof of Theorem 2 is deferred to Appendix B. According to Theorem 2, if we want to drive $\min_{t} \epsilon_t$ below a threshold $\epsilon > 0$, it suffices to take $T$ steps where 
\begin{equation}
	T \ge \frac{L_UR_U^2}{\epsilon}.
\end{equation}

\subsection{Optimal Distributed Routing Algorithm}
\label{footnote2}
Algorithm 1 needs to invoke an oracle $\mathfrak{O}$ to obtain the optimal routing decision $\bm{\phi}^{*}(\Lambda^{t})$ to minimize the total network cost $\sum_{(i,j) \in \bar{\mathcal{E}}}D_{ij}$ when given the workload allocation $\Lambda^{t}$. In this subsection, we try to find out such an oracle. Specifically, we define the optimal routing problem $\mathcal{P}2$ as follows:
\begin{align}
	\mathcal{P}2: \quad \min_{\bm{\phi}} \quad & D(\Lambda^{t},\bm{\phi}) \triangleq \sum_{(i,j) \in \bar{\mathcal{E}}}D_{ij}(F_{ij}(\bm{\phi}),C_{ij}) \label{r}\\
	\mathrm{s.t.} \quad 
	&\mathrm{\text{flow conservation constraints }} \eqref{a}-\eqref{d} \notag, \\
	&\mathrm{\text{and specially}}\sum_{j \in \mathcal{O}(i)} f_{ij}(w) = \lambda_w^{t}, \quad i = S,
\end{align}
where $F_{ij}(\bm{\phi})$ is the total flow rate on link $(i,j)$ when the routing decision is $\bm{\phi}$. We should note that the optimal routing problem $\mathcal{P}2$ is similar to the work \cite{xi2008node}. However, the proof of the optimality of $\mathcal{P}2$ is involved in \cite{xi2008node}, and the proposed routing algorithm has no convergence rate guarantee. In contrast, in this work, we provide a straightforward proof of the optimality that is only based on the convexity of $\mathcal{P}2$ and KKT conditions.  Moreover, we propose the mirror descent technique to update the routing variables, which enables that our routing algorithm exhibits the same convergence speed but requires less computation and communication overhead than its counterpart in \cite{xi2008node}. We further derive the convergence rate guarantee of our proposed routing algorithm.

We first give the theorem of the optimality conditions, which simplifies the Theorem 1 of \cite{xi2008node}. The proof of Theorem 3 is deferred to Appendix C.

\begin{mythm}
	Problem $\mathcal{P}2$ has a unique solution $\bm{\phi}^{*}(\Lambda^{t})$.  Moreover, the necessary and sufficient condition of optimality is for all $w \in \mathcal{W}$ and $i \notin D(w)$ with $t_i(w) > 0$, the following equation holds:
	\begin{equation}
		\frac{\partial D}{\partial \phi_{ij}^{*}(w)}  = -\alpha_i^{*}(w), \quad \forall j \in \mathcal{O}(i), \label{u}
	\end{equation}
	where $\alpha_i^{*}(w)$ is the optimal Lagrangian multiplier of the constraint $\sum_{j \in \mathcal{O}(i)} \phi_{ij}(w) = 1$.
\end{mythm}

\begin{myrem}
	The proof of Theorem 3 is similar to Theorem 1 where we first show the convexity of the optimal routing problem $\mathcal{P}2$. Then the derivation of the necessary and sufficient condition is based on the KKT optimal conditions. Finally, the equation \eqref{u} enlighten us an efficient distributed algorithm to find the optimal point $\bm{\phi}^{*}$ to minimize the problem $\mathcal{P}2$.
\end{myrem}

To clarify the optimality condition of the routing problem, it is necessary to derive the cost gradients with respect to the routing variables. Our analysis follows \cite{1093711}, specially, the gradient of $\phi_{ij}(w)$ is given by 
\begin{equation}
	\frac{\partial D}{\partial \phi_{ij}(w)} = t_i(w) \cdot \delta\phi_{ij}(w), \quad \forall j \in \mathcal{O}(i),
\end{equation}
where the marginal routing cost is 
\begin{equation}
	\delta \phi_{ij}(w) \triangleq \frac{\partial D_{ij}}{\partial F_{ij}} + \frac{\partial D}{\partial r_j(w)}. \label{ac}
\end{equation}
Here, the term $\frac{\partial D}{\partial r_j(w)}$ represents the marginal cost due to a unit increment of session $w$'s input rate at node $j$. It can be computed recursively by \cite{1093711}
\begin{align}
	\frac{\partial D}{\partial r_j(w)} &= 0, \quad \mathrm{\text{if }} j = D_w, \\ 
	\frac{\partial D}{\partial r_i(w)} &= \sum_{j \in \mathcal{O}(i)} \phi_{ij}(w)\Bigg[\frac{\partial D_{ij}}{\partial F_{ij}} + \frac{\partial D}{\partial r_j(w)}\Bigg] \notag \\
	&= \sum_{j \in \mathcal{O}(i)} \phi_{ij}(w) \cdot \delta \phi_{ij}(w), \quad \forall i \neq D_w. \label{ad}
\end{align}
We can see that the marginal cost for a marginal increase of $\phi_{ij}(w)$ consists of two components, 1) the direct marginal communication cost on link $(i,j)$, and 2) the indirect marginal cost of increasing input rate on downstream node $j$, which again is the sum of all marginal costs for $j$'s out-going links. Since $\frac{\partial D}{\partial \phi_{ij}(w)} = 0$ when $t_i(w)=0$ whatever $\delta\phi_{ij}(w)$ is, the marginal cost information is eliminated. Therefore, we only consider the nodes with $t_i(w) > 0$ in the rest of this paper.

Given the marginal costs $\delta\phi_{ij}(w)$, each distributed node can update its own routing decisions to collaborate to minimize the total network cost $D$, and each node may use different routing algorithms since they are autonomous. The biggest challenge is how to guarantee such distributed algorithm can achieve optimality because all marginal costs $\delta\phi_{ij}(w)$ are intersected and each node makes decision only based on its own information.

Since $D$ is convex on $\bm{\phi}$ and according to the optimality conditions \eqref{u}, the class of scaled gradient projection algorithms is suitable for providing a distributed solution \cite{1093711,bertsekas1997nonlinear}. To the best of our research, \cite{xi2008node} is the state-of-the-art of the class of  gradient projection algorithm (SGP) which exhibits the best convergence performance and communication overhead.

In this paper, we further promote the family of SGP algorithms by exploiting the online mirror descent (OMD) technique. Specially, the OMD takes the topology of the routing decision space into consideration, which exhibits much faster convergence rate than the traditional gradient descent and is comparable to the second order Newton method. Furthermore, we show that our proposed OMD based routing algorithm enjoys the same convergence rate with the SGP algorithms while requiring less computation and communication overhead\footnote{The scaling matrix $M_i^{k}(w)$ in \cite{xi2008node} is carefully chosen for upper bounding the Hessian which leads to larger stepsize, and therefore faster convergence rate. However, the computation of $M_i^{k}(w)$ needs many extra information such as the initial total cost at  $\bm{\phi}^{1}$, the maximum path length from node $i$ to destination $D_w$, \emph{et al}, just name a few. However, our proposed algorithm does not need such extra system information.}. The proposed method is summarized in Algorithm 2.\footnote{The Algorithm 2 is used as the oracle algotirhm $\mathfrak{O}$ to obtain the optimal routing decision $\bm{\phi}^{*}(\Lambda^{t})$ when the workload allocation is $\Lambda^{t}$.}

\begin{algorithm}[t] 
	\caption{The optimal routing algorithm OMD-RT} 
	\label{alg2} 
	\begin{algorithmic}[1] 
		\REQUIRE  
		The workload allocation $\Lambda^{t}$, the stepsizes $\eta_k$
		\ENSURE 
		The optimal routing decision $\bm{\phi}^{*}$
		\STATE \textbf{Initialize:} $\bm{\phi}_i^{1}(w) = \frac{1}{|\mathcal{O}(i)|} \cdot \mathbf{1}$, where $\mathbf{1}$ is the all one vector with coordinate $|\mathcal{O}(i)|$
		\STATE \textbf{for} each inner loop $k = 1,\ldots, K$ \textbf{do}
		\STATE \quad perform broadcast to obtain $\frac{\partial D}{\partial r_i(w)}$
		\STATE \quad calculate $\delta\phi_{ij}(w)$ according to \eqref{ac}
		\STATE \quad \textbf{for} all $i \notin D(w)$ with $t_i(w) > 0$ \textbf{do}
		\begin{equation}
			\phi_{ij}^{k+1}(w) = \frac{\phi_{ij}^{k}(w)\exp(-\eta_k \delta\phi_{ij}(w))}{\sum_{j}\phi_{ij}^{k}(w)\exp(-\eta_k \delta\phi_{ij}(w))} \label{ae}
		\end{equation}
		\STATE \quad \textbf{if} $\bm{\phi}_i^{k+1}(w)==\bm{\phi}_i^{k}(w)$, \textbf{break}
		\STATE \textbf{end for}
	\end{algorithmic}
\end{algorithm} 
\textbf{Marginal cost broadcast.} Each node $i$ needs to calculate the gradient $\delta\phi_{ij}(w)$ following \eqref{ac}. As the closed-form of $D_{ij}$ are known, nodes can directly calculate the value of $\frac{\partial D_{ij}}{\partial F_{ij}}$ while sending workloads on link $(i,j)$. To recursively obtain $\frac{\partial D}{\partial r_i(w)}$ from \eqref{ad}, we introduce a broadcast protocol: the broadcast of $\frac{\partial D}{\partial r_i(w)}$ starts with the last node of each path destined to $D_w$, and then provides its own marginal cost $\frac{\partial D}{\partial r_i(w)}$ to its neighbors from which it receives workload of session $w$. Since the routing variables are loop-free according to \eqref{a}-\eqref{d}, the broadcast processes are guaranteed to end within a finite number of steps.\footnote{We should note that the transmission cost of the marginal cost information is negligible since it is only a few bits that can be piggybacked by the task message transmitting between the adjacent nodes. As shown in Fig. 2, the largest path length equals to $W$, that is, the number of the DNN version. In practice, the number of DNN model version is quiet small,  for example, the popular video analyst model YOLOv5 has at most several versions for various settings, such as resolution and model size. Therefore, the broadcast procedure will finish in a few steps.}
\begin{myrem}
	Note that the exponentiated gradient descent \eqref{ae} in OMD-RT prefers routing more workload $\phi_{ij}(w)$ to the link that has smaller marginal cost $\delta\phi_{ij}(w)$. When the optimal condition \eqref{u} is achieved at $\bm{\phi}_i^{k}(w)$, we can immediately get that $\bm{\phi}_i^{k+1}(w)==\bm{\phi}_i^{k}(w)$ according to \eqref{ae}. 
\end{myrem}

We should point that, although the mirror descent is cataloged to the first-order technique, it exhibits a much faster convergence rate than the traditional gradient descent since the mirror descent fits the geometry of the feasible set $\mathcal{H}(\bm{\phi}_i(w))$.  Finally, we give the convergence rate guarantee of OMD-RT to finish the optimal routing subsection.
\begin{mythm}
	Assume that the optimal routing problem $D$ has $L_D$-Lipschitz continuous gradient on $\bm{\phi}$ for any given $\Lambda^{t}$, that is 
	\begin{equation}
		D(\tilde{\bm{\phi}}) \le D(\bm{\phi}) + \langle \frac{\partial D}{\partial \bm{\phi}}, \tilde{\bm{\phi}} - \bm{\phi} \rangle + \frac{L_D}{2}|| \tilde{\bm{\phi}} - \bm{\phi} ||^2, \quad \forall \tilde{\bm{\phi}}, \bm{\phi} \in \mathcal{H}(\bm{\phi}),
	\end{equation} 
	the online mirror descent based optimal routing algorithm OMD-RT converges with 
	\begin{equation}
		\min_{k \in \{2,\ldots,K+1\}} \epsilon_k \le \frac{L_DR_D^2}{cK},
	\end{equation}
	where $\epsilon_k \triangleq D(\bm{\phi}^{k}) - D(\bm{\phi}^{*})$ is the optimization error in inner loop $k$, and $R_D$ is the diameter of the decision space $\mathcal{H}(\bm{\phi})$.
\end{mythm}
The proof of Theorem 4 is deferred to Appendix D. Combining the Theorem 2 and 4, we can approximately bound the convergence rate for achieving the optimal point $(\Lambda^{*},\bm{\phi}^{*}(\Lambda^{*}))$ by $O(\frac{1}{TK})$, where $T$ and $K$ are the iteration numbers of outer and inner loop, respectively\footnote{We should note that our proposed method is similar in spirit
	with ADMM since both alternatively optimize different variable blocks. However, in the variable update iteration of ADMM, it allows slackness of the constraint of the two variable blocks, that is, the flow conservation constraints (1)-(4) may not be satisfied in every iteration step, while in
	our case these constraints are required to be satisfied.}. However, in a large network, the frequent node activity would change the network topology, and therefore the optimal routing configuration changes. And in the nested-loop algorithm, each outer loop needs to wait the inner loop to converge, which is clumsy to the change of the network. To tackle this weakness, we now explore the feasibility of the single-loop algorithm to well adapt the frequent change in the network.

\subsection{Improving Convergence Rate for JOWR}
In this subsection, we devise a single-loop algorithm to improve the convergence rate over the nested-loop version introduced in previous two subsections. According to Theorem 1 and 3, we know that the function $U(\Lambda,\bm{\phi})$ is concave-convex, and there exists one and only one optimal point $(\Lambda^{*},\bm{\phi}^{*}(\Lambda^{*}))$ that satisfies
\begin{equation}
	U(\Lambda,\bm{\phi}^{*}(\Lambda))  \le U(\Lambda^{*},\bm{\phi}^{*}(\Lambda^{*}))  \le
	U(\Lambda^{*},\bm{\phi}(\Lambda^{*})).
\end{equation}
Fortunately, there are numerous works to study algorithms and their convergence in solving this max-min problem, especially in the context of concave-convex settings \cite{pmlr-v139-yoon21d} \cite{doi:10.1137/19M127375X}. Inspired by the method in \cite{doi:10.1137/19M127375X} that the outer loop updates do not need to wait until the inner loop to converge, we propose our online mirror ascent descent (OMAD) single-loop algorithm for JOWR in Algorithm 3. Note that different from the method in \cite{doi:10.1137/19M127375X} that needs to compute the momentum term or extra-gradient information to perform gradient iteration, our proposed OMAD only needs to perform once mirror ascent or descent in each iteration which is more lightweight. Moreover, the mirror ascent and descent technique can well fit the geometry of decision variables in problem JOWR which usually exhibits faster convergence.

\begin{algorithm}[t] 
	\caption{The online mirror ascent descent (OMAD) single-loop algorithm for JOWR} 
	\label{alg3} 
	\begin{algorithmic}[1] 
		\REQUIRE  
		The disturbance $\delta$, the step sizes $\eta_o$ and $\eta_i$ for \eqref{s} and \eqref{ae}
		\ENSURE 
		The optimal point $(\Lambda^{*}, \bm{\phi}^{*}(\Lambda^{*}))$
		\STATE \textbf{Initialize:} $\Lambda^1=\frac{\lambda}{W} \cdot \mathbf{1}$
		\STATE \textbf{for} each single-loop $t = 1,\ldots, T$ \textbf{do}
		\STATE \quad \textbf{for} each session $w = 1,\ldots,W$ \textbf{do} 
		\STATE \quad \quad allocate $\Lambda^{+}(t) = \Lambda^{t} + \delta \bm{e}_w$ and invoke Algorithm 2 with 
		
		\quad \quad $K=1$ to observe $U^{+} = U(\Lambda^{+}(t),\tilde{\bm{\phi}}(\Lambda^{+}(t)))$
		\STATE \quad \quad allocate $\Lambda^{-}(t) = \Lambda^{t} - \delta \bm{e}_w$ and invoke Algorithm 2 with 
		
		\quad \quad $K=1$ to observe $U^{-} = U(\Lambda^{-}(t),\tilde{\bm{\phi}}(\Lambda^{-}(t)))$
		\STATE \quad \quad gradient estimate: $\frac{\partial U}{\partial \lambda_w^{t}}=\frac{U^{+}-U^{-}}{2\delta}$
		\STATE \quad \textbf{end for}
		\STATE \quad workload allocation update using \eqref{s}
		\STATE \quad projection step: $\Lambda^{t+1} = \mathcal{P}_{[\delta, \lambda-\delta]^{W}}[\Lambda^{t+1}]$
		\STATE \quad \textbf{if} $\Lambda^{t+1}==\Lambda^{t}$, \textbf{break}
		\STATE \textbf{end for}
	\end{algorithmic}
\end{algorithm}

The key distinction between Algorithms 1 and 3 is that the single-loop algorithm only needs to execute once the routing variables are iterated to improve the routing decision $\tilde{\bm{\phi}}$. In contrast, the nested-loop algorithm necessitates waiting until the routing variables converge to $\bm{\phi}^{*}$. Regarding the convergence of the single-loop algorithm, we can provide the following guarantee:

\begin{mythm}
	The single-loop Algorithm 3 solves the problem \eqref{g} at a rate $O(\frac{1}{t})$.
\end{mythm}
The proof of Theorem 5 is deferred to Appendix E. Note that our convergence rate proof is based on Lyapunov functions which is more straightforward compared to \cite{doi:10.1137/19M127375X}.

\section{Performance Evaluation}
\label{performance eval}
\begin{figure*}[t]
	\centering
	\begin{minipage}[t]{0.24\linewidth}
		\centering
		\includegraphics[width=1.5in]{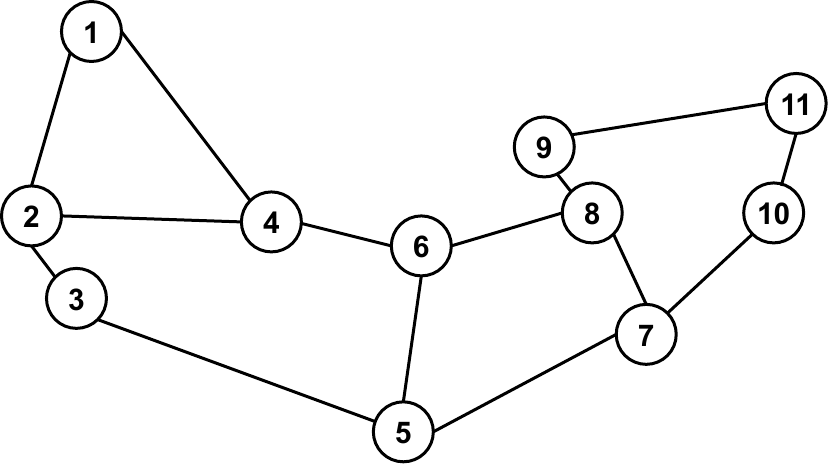}
		\caption{Abilene Topology.}
	\end{minipage}%
	\begin{minipage}[t]{0.24\linewidth}
		\centering
		\includegraphics[width=1.5in]{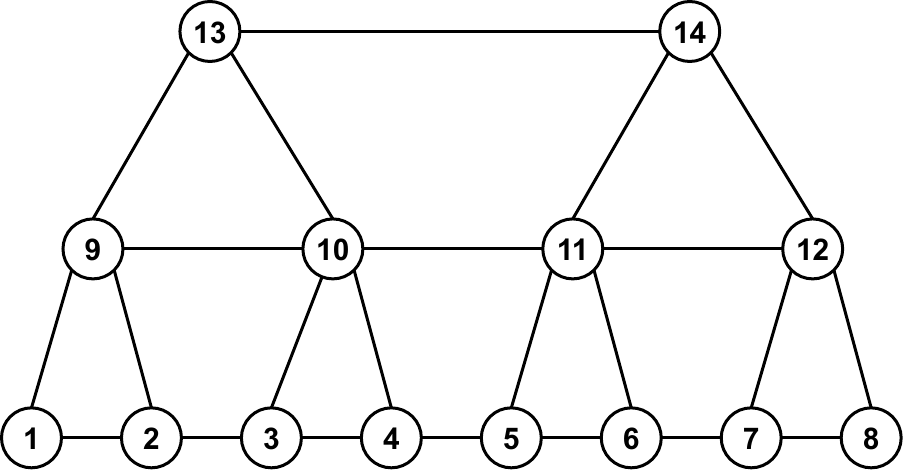}
		\caption{Tree Topology.}
	\end{minipage}%
	\begin{minipage}[t]{0.24\linewidth}
		\centering
		\includegraphics[width=1.3in]{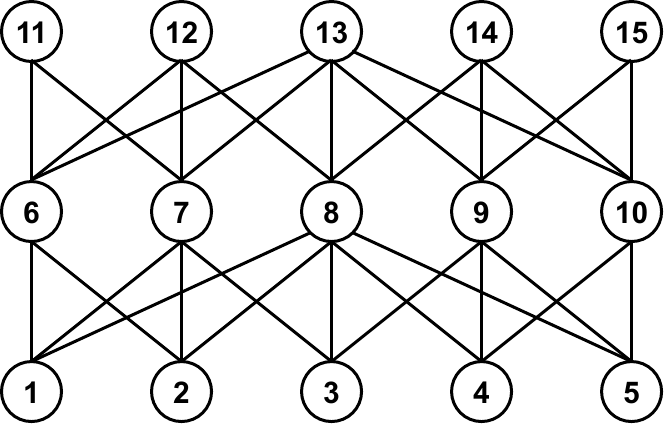}
		\caption{Fog Topology.}
	\end{minipage}
	\begin{minipage}[t]{0.24\linewidth}
		\centering
		\includegraphics[width=1.2in]{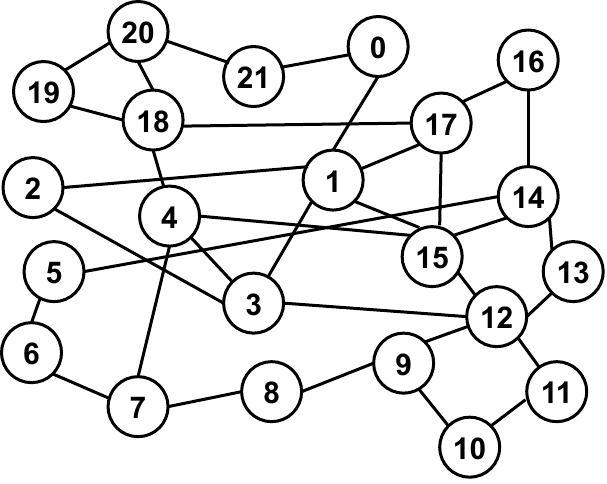}
		\caption{GEANT Topology.}
	\end{minipage}
\end{figure*}

	\begin{figure*}[t]
	\centering
	\begin{minipage}[t]{0.33\linewidth}
		\centering
		\includegraphics[width=2.2in]{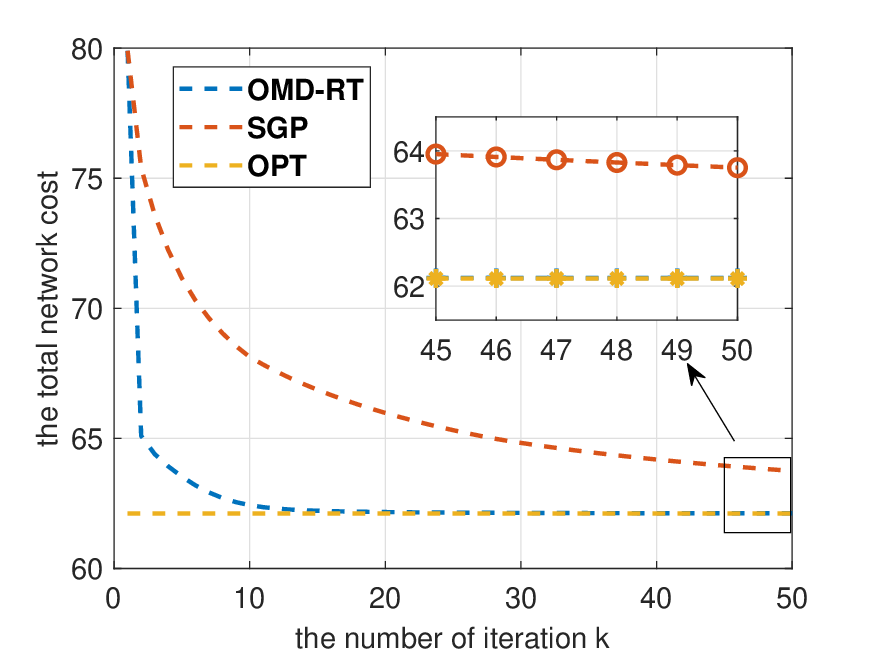}
		\caption{Convergence performance of the \\optimal routing algorithm OMD-RT.}
	\end{minipage}%
	\begin{minipage}[t]{0.33\linewidth}
		\centering
		\includegraphics[width=2.1in]{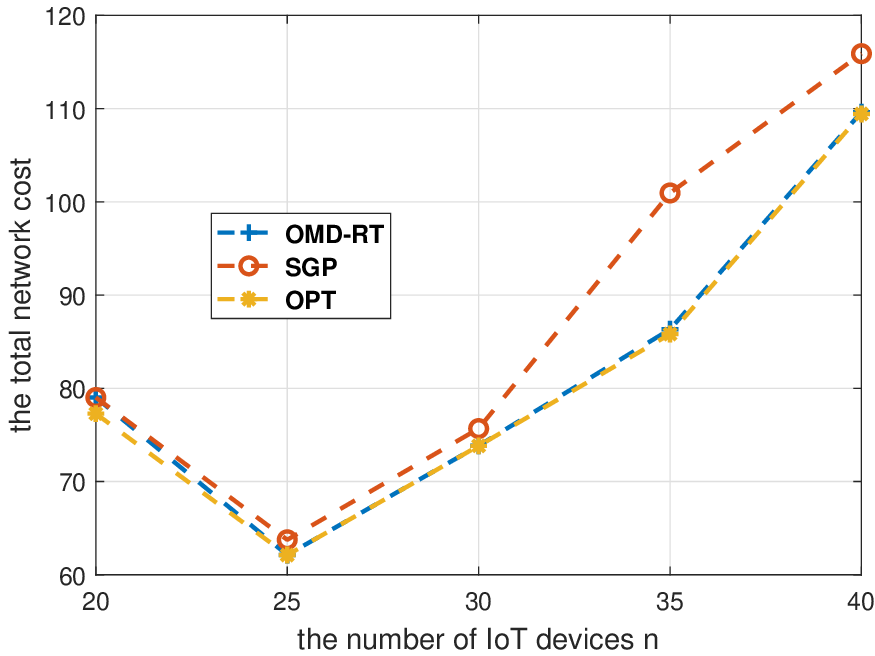}
		\caption{The total network cost under \\different network size.}
	\end{minipage}%
	\begin{minipage}[t]{0.33\linewidth}
		\centering
		\includegraphics[width=2.2in]{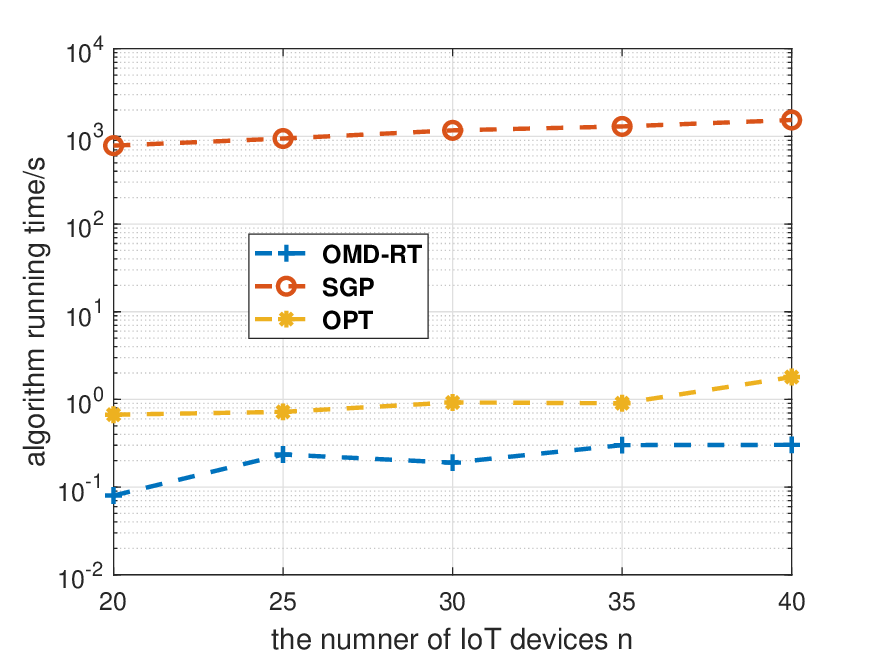}
		\caption{The algorithm running time under \\different network size.}
	\end{minipage}%
\end{figure*}

This section demonstrates our experimental evaluation of the proposed algorithms. First, we compare our optimal routing algorithm OMD-RT with the state-of-the-art SGP algorithm on different network topologies: \textbf{Connected-ER}$(n,p)$ is a connectivity-guaranteed Erdos-Renyi graph that generated by uniformly-randomly creating links with probability $p$ among $n$ nodes. \textbf{Abilene} (Figure 3) is the topology of the predecessor of \emph{Internet2 Network} \cite{rossi2011caching}. \textbf{Balance-tree} (Figure 4) is a complete tree. \textbf{Fog} (Figure 5) is a sample topology for fog-computing \cite{kamran2019deco}. \textbf{GEANT} (Figure 6) is a pan-European data network for the research and education community \cite{rossi2011caching}. Next, we assess the performance of the nested-loop Algorithm 1 for solving the JOWR problem under different kinds of the unknown utility functions. Finally, we test the convergence improvement of the single-loop Algorithm 3 in both static and changing networks.

\textbf{Experiment Setup.} In the Connected-ER topology, we assume there are  $n=25$ edge devices and each pair of them are connected with the probability $p=0.2$. The total DNN inference task input rate $\lambda$ is 60. The number of DNN model version $|\mathcal{W}|$ is 3, and each device is randomly deployed with one DNN version among the three.\footnote{Such settings can be instantiated as intelligent video resolution enhancement task with input rate 60 frames per second, and served by three output-resolution versions of DNN models (720P, 1080P or 2K)\cite{bishop2003super}. We should note that our proposed algorithms can well adapt to other parameter settings.} The link capacities $C_{ij}$ are uniformly drawn from $[0, 2\bar{C}_{ij}]$ with mean $\bar{C}_{ij}=10$ which follows \cite{zhang2022optimal}. For all experiments, we adopt $D_{ij}(F_{ij},C_{ij})=\exp(F_{ij}/C_{ij})$ as the link cost function. We only present the simulation results of the Connected-ER topology since the results of other topologies are similar. The detailed parameters setup  and simulation results of the rest topologies can be found in the Appendix F. 

Figure 7 shows the convergence performance of our proposed online mirror descent based optimal routing algorithm OMD-RT. We compare it with two benchmarks: \textbf{SGP}, to the best of our research, the scaled gradient projection algorithm is the state-of-the-art distributed optimal routing algorithm in wireless network with convex link cost. \textbf{OPT}, is the centralized optimal routing decision where the operator has access to the whole network topology, and then figures out all possible routing path from the source node to the destination node and finally solves the optimal routing via a convex solver. As shown in Figure 7, both OMD-RT and SGP converge to the optimal total network cost and therefore the optimal routing decisions. However, our proposed OMD-RT converges much faster than SGP in the first 10 iterations. This much faster convergence performance of OMD-RT in the first few iterations also presents in other network topology (can be found in the additional simulation result in Appendix). The zoom-in figure shows that after 50 routing iterations, OMD-RT almost approach the optimal while SGP still suffers a slow convergence rate. 

Figure 8 and 9 further verifies the strengths of the proposed OMD-RT. We change the number of nodes $n=[20,25,30,35,40]$ in the Connected-ER graph and both OMD-RT and SGP run 50 routing iterations. The dotted lines show the total network cost under different network sizes. We can see that OMD-RT always approaches the optimal within 50 routing iterations while the convergence of SGP may be influenced by the network size. One interesting discovery is that the total network cost is minimized at $n=25$. Intuitively, the network with more devices can utilize more communication and computation resources and therefore result in smaller network cost. One possible interpretation is that the performance of a collaborative network system not only depends on the total resources, but also depends on the network topology and the placement configuration of the DNN models. This is left for our future research. The solid lines show the algorithm running time. We can see that the computation of OMD-RT is significantly lightweight compared to SGP, at around three orders of magnitude running time improvement. Moreover, the running time of OMD-RT is shorter than OPT although it needs 50 routing iterations to converge. The reason is that both SGP and OPT need to solve a complex convex problem while OMD-RT just needs to execute a soft-max operation in each routing iteration.

	\begin{figure*}[t]
		\centering
		\begin{minipage}[t]{0.49\linewidth}
			\centering
			\includegraphics[width=2.4in]{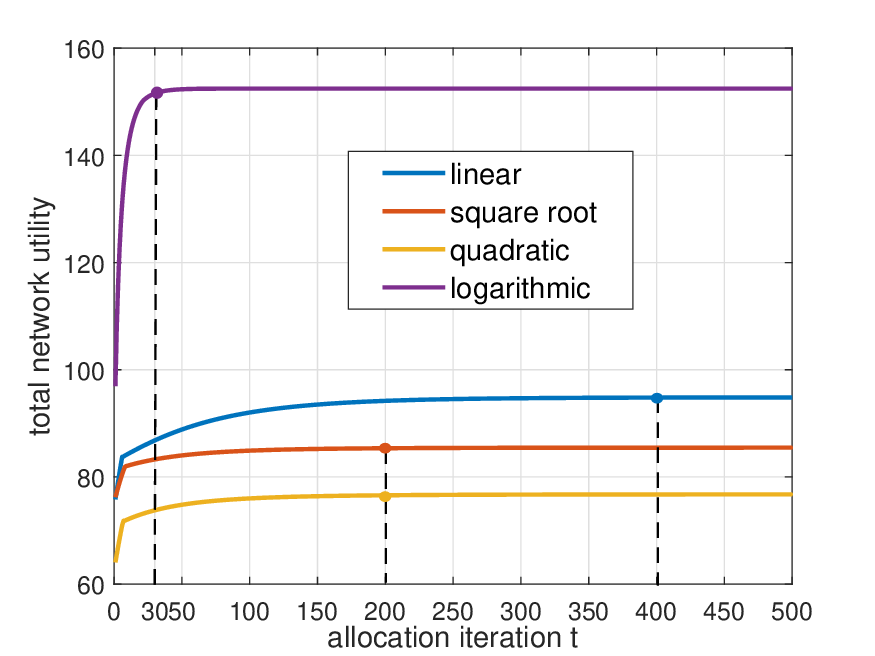}
			\caption{The total network utility of different kinds of \\utility functions.}
		\end{minipage}%
		\begin{minipage}[t]{0.49\linewidth}
			\centering
			\includegraphics[width=2.4in]{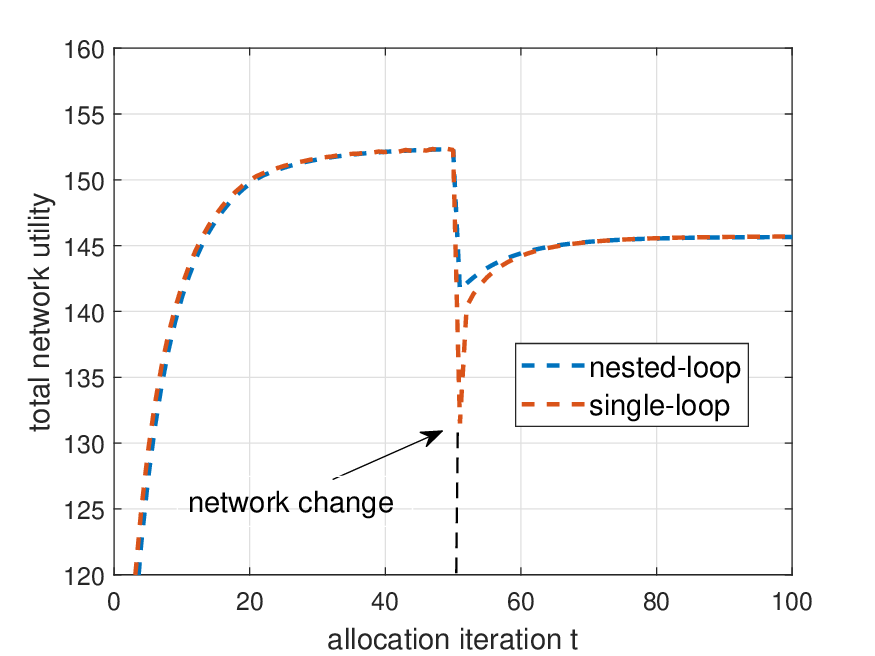}
			\caption{The total network utility of nested-loop and \\single-loop algorithms.}
		\end{minipage}%
	\end{figure*}

Next, we assess the performance of the nested-loop Algorithm 1 for solving the problem JOWR under four kinds of the unknown utility functions: $u_w(\lambda_{w})=a_w\lambda_w$ (linear function), $u_w(\lambda_{w})=a_w\sqrt{\lambda_w+b_w}-a_w\sqrt{\lambda_w}$ (square root function), $u_w(\lambda_{w})=-a_w\lambda_w^2+b_w\lambda_w$ (quadratic function), $u_w(\lambda_{w})=a_w\log(b_w\lambda_w+1)$ (logarithmic function). Since the settings of pair $(a_w,b_w)$ are different, the total network utilities converge to different values. As shown in figure 10, although we do not know the exact utility function when we need to admit the workload, using the gradient sampling technique to estimated gradient and then the online mirror ascent technique can converge to the optimal allocation configuration under different kinds of utility functions. Specially, the linear function needs around 400 iterations to converge while the logarithmic function needs around 30 iterations. This phenomenon shows that the gradient sampling technique may perform differently under distinct function curves. We use the logarithmic function as the utility function in the rest simulations.

Finally, the figure 11 verifies the convergence improvement of the single-loop Algorithm 3. Although the single-loop algorithm only performs once routing iteration in each outer workload allocation iteration, both the nested-loop and single-loop algorithm converge to the optimal point $(\Lambda^{*},\bm{\phi}^{*}(\Lambda^{*}))$. This phenomenon shows that in each workload allocation iteration, we do not need to wait the routing decisions to converge, which is a huge convergence time improvement for solving the JOWR problem. We further change the network topology at the 50-th allocation iteration, it can be seen that both nested-loop and single-loop algorithms can quickly adapt to the new network topology while the single-loop algorithm always starts at a worse initial point since the routing decisions are not optimal in the initial workload allocation iteration. 

\section{Related work}
Many previous studies \cite{zhang2022optimal, liu2020distributed, 9623540, sahni2018data} have primarily focused on the workload routing in CEC, assuming fixed workload allocation in advance. Some works consider edge devices in the CEC network, whether served as either relay nodes, responsible for transmitting data of computing tasks \cite{liu2020distributed, sahni2018data}, or processing nodes, computing partial workloads of the tasks \cite{9623540, zhang2022optimal}. Both scenarios require an optimal routing to minimize total communication and computation costs in the edge network. However, only a few works focus on joint optimization of workload allocation and routing in CEC \cite{neely2008fairness, xi2008node, jingzong2023cross}, aiming to control task input rates to maximize task utility. Nevertheless, all of these works consider known utility functions while the functional relationships between the task input rates and corresponding task utility values are usually unknown for more complicated applications in CEC. To the best of our knowledge, our work is the first to consider the joint optimization of workload allocation and routing under unknown utility function in resource-constrained CEC environment.

The work \cite{1093711} first adopted the gradient projection algorithm for distributed optimal routing in wireline network. However, the proposed algorithm suffered from a slow convergence rate due to its small step size. Subsequently, some researches resorted to the second order derivatives to improve the convergence rate of the gradient projection algorithms, leading to the development of the scaled gradient projection algorithm (SGP). As a variant of the Newton projection algorithm, SGP demonstrates super-linear convergence when the initial point is in close proximity to the optimum \cite{bertsekas1997nonlinear}. Nevertheless, the computational complexity associated with computing the exact Hessian matrix for the Newton method restricts its applicability to large-scale networks. To address the above problem, the work \cite{bertsekas1984second} proposed a diagonal approximation to the Hessian, which is utilized to scale the descent direction. Furthermore, the work \cite{xi2008node} have introduced the state-of-the-art the class of SGP algorithm, demonstrating superior convergence performance and reduced communication overhead compared to previous approaches. In this study, we further enhance the SGP algorithm family by incorporating online mirror descent (OMD), which achieves the same convergence rate as SGP algorithms while requiring lower computational and communication overhead. 

The most related work is \cite{zhang2022optimal} since its model and problem formulation share similarity with ours to some extent. However, our work offers a more direct and appropriate framework for jointly optimizing workload allocation and routing decisions in the context of DNN inference. Additionally, the cross-layer structure further inspires us to design the nested-loop algorithm, and thus achieving the optimal total network utility with faster convergence. Although work \cite{zhang2022optimal} can also consider the network utility by redefining the computation cost function, the computation of the marginal cost in work \cite{zhang2022optimal} still requires a known utility function. Our work considers a more general unknown utility function scenario and employ the gradient sampling technique in workload allocation optimization, and further provide the theoretical-guarantee convergence of GS-OMA. Furthermore, the SGP algorithm in work \cite{zhang2022optimal} needs to compute the scaling
matrix and then solve a quadratic programming to derive the optimal routing decision, which needs additional system information and heavy computation. In contrast, our proposed mirror decent algorithm exhibits less information requirement and compute light.

\section{Conclusion}
\label{conc}
In this paper, we formulate the joint optimal workload allocation and routing problem in collaborative edge computing as a network utility maximization problem. After deriving the sufficient and necessary conditions for optimality, we first propose a nested-loop algorithm to converge to the optimal allocation and routing decisions. At each outer loop we apply the gradient sampling technique and online mirror ascent to iterate the workload allocation decisions to maximize the workload utility under the unknown utility functions. At each inner loop, given a fixed workload allocation decision, we proposed a novel optimal distributed routing algorithm based on the online mirror descent algorithm to iterate the routing decisions. In order to improve the convergence rate we further proposed a single-loop algorithm where the allocation and routing decisions iterate at the same iteration. Both theoretical and extensive numerical simulations demonstrate the superior performance of our proposed algorithms.

\section{Appendix}
\subsection{The proof of Theorem 1} \label{appendix A}
Recall the optimal workload allocation problem $\mathcal{P}1$:
\begin{align}
	\max_{\Lambda} \quad U(\Lambda, \bm{\phi}^{*}(\Lambda)) &\triangleq  \sum_{w \in \mathcal{W}}u_w(\lambda_w) - \sum_{(i,j) \in \bar{\mathcal{E}}}D_{ij}(F_{ij}^{*}(\Lambda),C_{ij}) \label{25}\\
	\mathrm{s.t.} & \quad \forall \lambda_w \ge 0 \mathrm{\text{ and }}\sum_{w \in \mathcal{W}} \lambda_w = \lambda, \label{26}
\end{align}
	
We first prove that $U$ is concave on $\Lambda$. Obviously, the decision space \eqref{26} forms a simplex, which therefore is a convex set. Since the utility functions $u_w$ are all concave, the first term of the right hand side (RHS) of \eqref{25}, $\sum_{w \in \mathcal{W}}u_w(\lambda_w)$ is also concave on $\Lambda$. Specifically, one can verify the concavity of $\sum_{w \in \mathcal{W}}u_w(\lambda_w)$ by deriving the Hessian matrix on $\Lambda$. Since all $u_w$ are monotonically increasing, continuously differentiable and concave, the second derivatives of all $u_w$ are negative. Therefore, the Hessian matrix of $\sum_{w \in \mathcal{W}}u_w(\lambda_w)$ is a negative definite matrix, which shows the concavity of $\sum_{w \in \mathcal{W}}u_w(\lambda_w)$.

The last step of proving the concavity of $U$ is to show that the second term of RHS of \eqref{25}, $\sum_{(i,j) \in \bar{\mathcal{E}}}D_{ij}(F_{ij}^{*}(\Lambda),C_{ij})$ is convex on $\Lambda$. Although the hidden relation $F_{ij}^{*}(\Lambda)$ is unknown, we can still show that for each convex cost function $D_{ij}(\cdot, C_{ij})$, $F_{ij}^{*}(\Lambda)$ is a linear combination of all the elements of $\Lambda$, which means that $F_{ij}^{*}(\Lambda)$ is affine on $\Lambda$. By definition, given a workload allocation $\Lambda$ and corresponding optimal routing strategy $\bm{\phi}^{*}(\Lambda)$, we have
\begin{equation}
	F_{ij}^{*} = \sum_{w \in \mathcal{W}} t_i^{*}(w) \phi_{ij}^{*}(w), \label{j}
\end{equation}
where
\begin{equation}
	t_i^{*}(w) = \sum_{k \in \mathcal{I}(i)} f_{ki}^{*}(w) = \sum_{k \in \mathcal{I}(i)} t_k^{*}(w)\phi_{ki}^{*}(w) \label{k}
\end{equation}
is the optimal input rate of session $w$ at node $i$. Combining \eqref{j} and \eqref{k} we have 
\begin{equation}
	F_{ij}^{*} = \sum_{w \in \mathcal{W}} \sum_{k \in \mathcal{I}(i)} t_k^{*}(w)\phi_{ki}^{*}(w) \phi_{ij}^{*}(w).
\end{equation}
Continue to unpack $t_k^{*}(w)$, we finally end at the virtual source node $S$ with $t_S(w) = \lambda_w$ by definition \eqref{a}. Therefore, we can derive that
\begin{equation}
	F_{ij}^{*} = \sum_{w \in \mathcal{W}} \cdots \sum_{l \in \mathcal{I}(k)} \sum_{k \in \mathcal{I}(i)} \lambda_w\cdots\phi_{lk}^{*}(w)\phi_{ki}^{*}(w) \phi_{ij}^{*}(w), \label{v}
\end{equation}
which shows that $F_{ij}^{*}$ is a linear combination of all the elements of $\Lambda$, and therefore $F_{ij}^{*}$ is affine on $\Lambda$. Moreover, even given any routing strategy $\bm{\phi}$, the resulting flow rate $F_{ij}^{\bm{\phi}}(\Lambda)$ is also affine on $\Lambda$.

Finally, since each $D_{ij}(\cdot, C_{ij})$ is an increasing, continuously differentiable and convex function for any fixed $C_{ij}$, the composition
\begin{equation}
	\sum_{(i,j) \in \bar{\mathcal{E}}}D_{ij}(F_{ij}^{*}(\Lambda),C_{ij}) 
\end{equation}
is still convex on $\Lambda$. Now, we have proven the first part of Theorem 1 that problem \eqref{i} is concave on $\Lambda$, therefore, problem \eqref{i} has one and only one optimal point $\Lambda^{*}$.

We proceed to prove the necessary and sufficient condition of optimality. The Lagrangian function of \eqref{i} can be presented formally as:
\begin{equation}
	L(\Lambda, \alpha, \bm{\beta}) = U - \alpha(\sum_{w \in \mathcal{W}} \lambda_w - \lambda) + \sum_{w \in \mathcal{W}}\beta_w\lambda_w,
\end{equation}
where $\alpha \ge 0$ and $\bm{\beta} \ge 0$ are the Lagrangian multipliers.
Recall that the KKT necessary and sufficient conditions are obtained by setting the derivatives of Lagrangian function to 0. Specifically, if $\Lambda^{*}$ optimally solves $\mathcal{P}1$, then there must exist multipliers $\alpha^{*}$ and $\bm{\beta}^{*}$ such that the following holds:
\begin{equation}
	\begin{aligned}
		&\frac{\partial U}{\partial \lambda_w^{*}} - \alpha^{*} + \beta_w^{*} = 0, \quad \forall w \in \mathcal{W} \\
		&\sum_{w \in \mathcal{W}} \lambda_w^{*} - \lambda = 0 \\
		&\lambda_w^{*} \ge 0, \quad \forall w \in \mathcal{W}\\ &\alpha^{*}(\sum_{w \in \mathcal{W}} \lambda_w^{*} - \lambda) = 0 \\
		&\beta_w^{*}\lambda_w^{*} = 0, \quad \forall w \in \mathcal{W}\\
		&\alpha^{*} \ge 0 \mathrm{\text{ and }} \beta_w^{*} \ge 0, \quad \forall w \in \mathcal{W}
		\end{aligned}
\end{equation}
Therefore, we can obtain that $\forall w \in \mathcal{W}$:
\begin{equation}
	\frac{\partial U}{\partial \lambda_w^{*}}=
	\begin{cases}
		 \alpha^{*} - \beta_w^{*}, & \mathrm{\text{ if }} \lambda_w^{*} = 0,\\
		 \alpha^{*}, & \mathrm{\text{ if }} \lambda_w^{*} > 0.
	\end{cases} \label{l}
\end{equation}
That is, in order to maximize $U$, the controller should only allocate workloads to the elements that have the maximum subgradient. The last step is to prove 
\begin{equation}
	\frac{\partial U}{\partial \lambda_1^{*}} = \cdots = \frac{\partial U}{\partial \lambda_w^{*}} = \cdots = \frac{\partial U}{\partial \lambda_W^{*}} = \alpha^{*}, \label{x}
\end{equation}
which means that all the elements in the optimal point $\Lambda^{*}$ should have the same subgradient. We prove \eqref{l} by a contradiction. Suppose there exist two optimal points $\Lambda_1^{*}$ and $\Lambda_2^{*}$ both satisfy \eqref{l}, then we can immediately get that $U(\Lambda_1^{*}) = U(\Lambda_2^{*})$. And there exists an element $\lambda_{1k}^{*} = 0$ in the optimal point $\Lambda_1^{*}$ with $\frac{\partial U}{\partial \lambda_{1k}^{*}} = \alpha^{*} - \beta_w^{*} < \alpha^{*} $, and the optimal point $\Lambda_2^{*}$ contains $\lambda_{2w}^{*} > 0, \forall w \in \mathcal{W}$.

Since $\Lambda_1^{*}$ and $\Lambda_2^{*}$ maximize $U$, we must have:
\begin{align}
	&U(\Lambda_2^{*}) \le U(\Lambda_1^{*}) + \langle \frac{\partial U}{\partial \Lambda_1^{*}}, \Lambda_2^{*} - \Lambda_1^{*} \rangle, \label{y}\\
	&U(\Lambda_1^{*}) \le U(\Lambda_2^{*}) + \langle \frac{\partial U}{\partial \Lambda_2^{*}}, \Lambda_1^{*} - \Lambda_2^{*} \rangle. \label{z}
\end{align}
It is easy to see that $\langle \frac{\partial U}{\partial \Lambda_2^{*}}, \Lambda_1^{*} - \Lambda_2^{*} \rangle = 0$ since $\sum_{w \in \mathcal{W}} (\lambda_{1w}^{*}-\lambda_{2w}^{*}) = 0$, and $\frac{\partial U}{\partial \lambda_{2w}^{*}} = \alpha^{*}, \forall w \in \mathcal{W}$. Therefore, the inequation \eqref{z} always holds, which shows that $\Lambda_2^{*}$ is the optimal point with all elements are positive. On the contrary, since $\Lambda_1^{*}$ has an element $\lambda_{1k}^{*} = 0$ with $\frac{\partial U}{\partial \lambda_{1k}^{*}} < \alpha^{*} $, we always have $\langle \frac{\partial U}{\partial \Lambda_1^{*}}, \Lambda_2^{*} - \Lambda_1^{*} \rangle < 0$, which is a contradiction with \eqref{y}. To sum up above, we have shown that the optimal point of problem \eqref{r} must satisfy \eqref{x}.

Furthermore, the readers can understand the \eqref{x} by intuition: since $U$ is concave on $\Lambda$, the optimal point $\Lambda^{*}$ must stay in the interior of the decision space $\mathcal{H}(\Lambda)$ which is a bowl-shape in high dimension space.

\subsection{The proof of Theorem 2} \label{appendix B}
For analysis simplicity, we set the total request rate $\lambda = 1$, therefore the decision space $\mathcal{H}(\Lambda)$ reduces to a probability simplex. In this setting, we want to maximize the concave function $U$ over a probability simplex $\mathcal{H}(\Lambda)$. Recall the subgradient mirror ascent rule
\begin{equation}
	\Lambda^{t+1} = \mathop{\arg\max}\limits_{\Lambda \in \mathcal{H}(\Lambda)} \Big\{ U(\Lambda^{t}) + \langle \bm{g}^t, \Lambda - \Lambda^{t}\rangle  - \frac{1}{\eta_t}\Delta_{\psi}(\Lambda, \Lambda^{t})
	\Big\}, \label{m}
\end{equation}
where $\bm{g}^t \in \partial U(\Lambda^{t})$ is the subgradient and $\Delta_{\psi}(\Lambda, \Lambda^{t})$ is the Bregman divergence
\begin{equation}
	\Delta_{\psi}(\Lambda, \Lambda^{t}) = \psi(\Lambda) - \psi(\Lambda^{t}) - \langle \nabla\psi(\Lambda^{t}), \Lambda - \Lambda^{t} \rangle
\end{equation}
with the measure function $$\psi(\Lambda)=\sum_{w \in \mathcal{W}} \lambda_w\log \lambda_w.$$ In such setting, the Bregman divergence reduces to the KL-divergence which measures the distance between two probability distribution $\Lambda$ and $\Lambda^{t}$. Specifically, the update rule \eqref{m} can be interpreted as follows. First approximate $U$ around $\Lambda^{t}$ by the first-order Taylor expansion 
\begin{equation}
	U(\Lambda) \approx U(\Lambda^{t}) + \langle \bm{g}^t, \Lambda - \Lambda^{t}\rangle.
\end{equation}
Then penalize the displacement by $-\frac{1}{\eta_t}\Delta_{\psi}(\Lambda, \Lambda^{t})$. The first order optimality condition for \eqref{m} is 
	\begin{align}
		&\bm{g}^t - \frac{1}{\eta_t}(\nabla\psi(\Lambda^{t+1}) - \nabla\psi(\Lambda^{t})) = 0 \label{n}\\
		\Longleftrightarrow \quad & \nabla\psi(\Lambda^{t+1}) = \nabla\psi(\Lambda^{t}) + \eta_t\bm{g}^t \notag\\
		\Longleftrightarrow \quad & \Lambda^{t+1} = (\nabla\psi)^{-1}(\nabla\psi(\Lambda^{t}) + \eta_t\bm{g}^t) \label{o}
	\end{align}
where \eqref{n} uses the property $\frac{\partial}{\partial x}\Delta_{\psi}(x,y) = \nabla\psi(x) - \nabla\psi(y)$. Therefore, the update rule of $\Lambda^{t+1}$ has the following closed form 
\begin{equation*}
	\lambda_w^{t+1} = \frac{\lambda_w(t)\exp(\eta_t\frac{\partial U}{\partial \lambda_w(t)})}{\sum_{w=1}^{W}\lambda_w(t)\exp(\eta_t\frac{\partial U}{\partial \lambda_w(t)})}, \quad \forall w \in \mathcal{W},
\end{equation*}
which is exactly the Line 8 of GS-OMA algorithm.

We now show that after every iteration, the total network utility would increase, that is 
\begin{equation}
	U(\Lambda^{t+1}) \ge U(\Lambda^{t}), \quad \forall t \in \{1,\ldots,T\}.
\end{equation}
Recall that $U$ has $L_U$-Lipschitz continuous gradient:
\begin{equation}
	U(\Lambda) \ge U(\Lambda^{t}) + \langle \bm{g}^t, \Lambda - \Lambda^{t}\rangle - \frac{L_U}{2}||\Lambda - \Lambda^{t}||^2.
\end{equation}
Since $\psi(\Lambda)$ is 1-strongly convex, we have 
\begin{equation}
	\Delta_{\psi}(\Lambda, \Lambda^{t}) \ge \frac{1}{2}||\Lambda - \Lambda^{t}||^2.
\end{equation}
Therefore, when we set $\eta_t \le \frac{1}{L_U}$, we can get
\begin{equation}
	\begin{aligned}
		U(\Lambda) &\ge U(\Lambda^{t}) + \langle \bm{g}^t, \Lambda - \Lambda^{t}\rangle - \frac{L_U}{2}||\Lambda - \Lambda^{t}||^2 \\
		&\ge U(\Lambda^{t}) + \langle \bm{g}^t, \Lambda - \Lambda^{t}\rangle - \frac{1}{2\eta_t}||\Lambda - \Lambda^{t}||^2 \\
		&\ge U(\Lambda^{t}) + \langle \bm{g}^t, \Lambda - \Lambda^{t}\rangle  - \frac{1}{\eta_t}\Delta_{\psi}(\Lambda, \Lambda^{t})
	\end{aligned}
\end{equation}
Finally, since $\Lambda^{t+1}$ is the maximizer of the right-hand-side of above inequation, we have
\begin{equation}
	\begin{aligned}
		&U(\Lambda^{t+1}) \ge U(\Lambda^{t}) + \langle \bm{g}^t, \Lambda^{t+1} - \Lambda^{t}\rangle  \\
		&\qquad \qquad \qquad - \frac{1}{\eta_t}\Delta_{\psi}(\Lambda^{t+1}, \Lambda^{t})\\
		\Longleftrightarrow & U(\Lambda^{t+1}) - U(\Lambda^{t}) \ge \langle \bm{g}^t, \Lambda^{t+1} - \Lambda^{t}\rangle  \\
		&\qquad \qquad \qquad \qquad \qquad \quad - \frac{1}{\eta_t}\Delta_{\psi}(\Lambda^{t+1}, \Lambda^{t})\\
		\Longleftrightarrow & U(\Lambda^{t+1}) - U(\Lambda^{t}) \ge 0.
	\end{aligned}	
\end{equation}
The last inequation is because the function $\langle \bm{g}^t, \Lambda - \Lambda^{t}\rangle  - \frac{1}{\eta_t}\Delta_{\psi}(\Lambda, \Lambda^{t})$ has a value 0 at point $\Lambda = \Lambda^{t}$, and $\Lambda^{t+1}$ is a maximizer of it, then we have $\langle \bm{g}^t, \Lambda^{t+1} - \Lambda^{t}\rangle  
 - \frac{1}{\eta_t}\Delta_{\psi}(\Lambda^{t+1}, \Lambda^{t}) \ge 0$

We continue to analyze the rate of convergence of subgradient mirror ascent in Algorithm 1. It takes four steps. \\
1. Bounding on a single update
\begin{equation}
\begin{aligned}
	\Delta_{\psi}(\Lambda^{*}, \Lambda^{t+1}) &\le \Delta_{\psi}(\Lambda^{*}, \Lambda^{t}) - \eta_t(U(\Lambda^{*}) - U(\Lambda^{t+1}) \\&- \frac{L_U}{2}||\Lambda^{t+1} -\Lambda^{t}||^2) 
	- \frac{1}{2}||\Lambda^{t+1} -\Lambda^{t}||^2 \label{p}
\end{aligned}	
\end{equation}
We omit the lengthy proof of \eqref{p} since this proof is not our main contribution. Readers are recommended to find the original proof in \cite{Bregman}. By setting $\eta_t=\frac{1}{L_U}$, we can get 
\begin{equation}
	\Delta_{\psi}(\Lambda^{*}, \Lambda^{t+1}) \le \Delta_{\psi}(\Lambda^{*}, \Lambda^{t}) - \frac{1}{L_U}(U(\Lambda^{*}) - U(\Lambda^{t+1})
\end{equation}\\
2. Telescope over $t = 1, \ldots, T$:
\begin{equation}
\Delta_{\psi}(\Lambda^{*}, \Lambda^{t+1}) \le \Delta_{\psi}(\Lambda^{*}, \Lambda(1)) - \sum_{t=1}^{T}\frac{1}{L_U}(U(\Lambda^{*})-U(\Lambda^{t+1})).
\end{equation}
3. Bounding the $\Delta_{\psi}(\Lambda^{*}, \Lambda(1))$ with $R_U^2$ and rearrange: 
\begin{equation}
	\sum_{t=2}^{T+1}\frac{1}{L_U}(U(\Lambda^{*})-U(\Lambda^{t})) \le R^{2} 
\end{equation}
By definition, $R_U^2 \triangleq \max_{\Lambda \in \mathcal{H}(\Lambda)} \Delta_{\psi}(\Lambda, \Lambda(1))$ is the diameter of the decision space $\mathcal{H}(\Lambda)$ with the measure function $\psi$.\\
4. Denote $\epsilon_t \triangleq U(\Lambda^{*}) - U(\Lambda^{t})$ and rearrange
\begin{equation}
	\min_{t \in \{2,\ldots,T+1\}} \epsilon_t \le \frac{L_UR_U^2}{T} \label{q}
\end{equation}
This gives $O(\frac{1}{T})$ convergence rate. Moreover, $R_U^2$ is at most $\log W$ if we set $\Lambda(1) = \frac{1}{W}\bm{1}$ and $\Lambda^{*}$ lies in the probability simplex $\mathcal{H}(\Lambda)$. Therefore, \eqref{q} can be rewritten as:
\begin{equation}
	\min_{t \in \{2,\ldots,T+1\}} \epsilon_t \le \frac{L_U \log W}{T}. \label{t}
\end{equation}

\subsection{The proof of Theorem 3} \label{appendix C}
Recall the optimal routing problem:
\begin{align}
	\min_{\bm{\phi}} \quad & D(\Lambda^{t},\bm{\phi}) \triangleq \sum_{(i,j) \in \bar{\mathcal{E}}}D_{ij}(F_{ij}(\bm{\phi}),C_{ij}) \\
	\mathrm{s.t.} \quad 
	&\mathrm{\text{flow conservation constraints }} \eqref{a}-\eqref{d} \notag, \\
	&\mathrm{\text{and specially}}\sum_{j \in \mathcal{O}(i)} f_{ij}(w) = \lambda_w(t), \quad i = S.
\end{align}

We first prove that $D$ is convex on $\bm{\phi}$, which is hard to recognize at first glance. To show that the decision space $\mathcal{H}(\bm{\phi})$ of routing variables $\bm{\phi}$ is a convex set, we can divide $\mathcal{H}(\bm{\phi})$ into a Cartesian product of a series of convex sets:
\begin{equation}
	\begin{aligned}
		\mathcal{H}(\bm{\phi}) &= \mathcal{H}(\bm{\phi}_i(1)) \times \cdots \times \mathcal{H}(\bm{\phi}_i(w)) \times \cdots 
		\times \mathcal{H}(\bm{\phi}_i(W)) \\ 
		&\times \cdots \\
		&\times \mathcal{H}(\bm{\phi}_j(1))  \times \cdots \times \mathcal{H}(\bm{\phi}_j(w)) \times \cdots 
		\times \mathcal{H}(\bm{\phi}_j(W)),
	\end{aligned}
\end{equation}
where $\mathcal{H}(\bm{\phi}_i(w))$ is the decision space of routing variables $\bm{\phi}_i(w) \triangleq \{\phi_{ij}(w):j \in \mathcal{O}(i)\}$ at node $i$ for session $w$. Obviously, $\mathcal{H}(\bm{\phi}_i(w))$ is a convex set according to flow conservation constraints \eqref{a}-\eqref{d}. Specifically, for all $w \in \mathcal{W}$ and $i \notin D(w)$ with $t_i(w) > 0$, the decision space $\mathcal{H}(\bm{\phi}_i(w))$ is a probability simplex according to the constraint $\sum_{j \in \mathcal{O}(i)} \phi_{ij}(w) = 1$.

Since $D$ is the sum of $D_{ij}(F_{ij}(\bm{\phi}),C_{ij})$, we only need to show that $D_{ij}(F_{ij}(\bm{\phi}),C_{ij})$ is convex on $\bm{\phi}$. According to \eqref{v}, we have
\begin{equation}
	F_{ij}(\bm{\phi}) = \sum_{w \in \mathcal{W}} \cdots \sum_{l \in \mathcal{I}(k)} \sum_{k \in \mathcal{I}(i)} \lambda_w(t)\cdots\phi_{lk}(w)\phi_{ki}(w)\phi_{ij}(w)
\end{equation}
For each addend $\lambda_w(t)\cdots\phi_{lk}(w)\phi_{ki}(w)\phi_{ij}(w)$, we can easily see that it is convex on $\bm{\phi}$. And $F_{ij}$ is the sum of all addends, therefore, we can conclude that $F_{ij}$ is also convex on $\bm{\phi}$. Together with $D_{ij}(\cdot, C_{ij})$ is a convex function for any fixed $C_{ij}$, we have that $D_{ij}$ is also convex on $\bm{\phi}$. Finally, the optimal routing problem $D$ is the sum of all $D_{ij}$ which therefore is a convex function on $\bm{\phi}$. Now, we have proven the first part of Theorem 3 that problem \eqref{r} is convex on $\bm{\phi}$, therefore, problem \eqref{r} has one and only one optimal point $\bm{\phi}^{*}$ that minimize the value $D$.

The proof of necessary and sufficient condition of optimality in Theorem 3 is similar to Theorem 1. We first give the Lagrangian function of problem \eqref{r} and only consider the routing variables for all $w \in \mathcal{W}$ and $i \notin D(w)$ with $t_i(w) > 0$:
\begin{equation}
	\begin{aligned}
		L(\bm{\phi}, \bm{\alpha}, \bm{\beta}) = D &+ \sum_{i \in \bar{\mathcal{N}}}\alpha_i(w)(\sum_{j \in \mathcal{O}(i)}\phi_{ij}(w) - 1) \\
		&- \sum \beta_{ij}(w)\phi_{ij}(w),
	\end{aligned}
\end{equation}
where $\bm{\alpha} \ge 0$ and $\bm{\beta} \ge 0$ are the Lagrangian multipliers for constrains $\sum_{j \in \mathcal{O}(i)} \phi_{ij}(w) = 1$ and $\phi_{ij}(w) \ge 0$, respectively. According to the KKT necessary and sufficient conditions, if $\bm{\phi}^{*}$ optimally minimize \eqref{r}, then there must exist multipliers $\bm{\alpha}^{*}$ and $\bm{\beta}^{*}$ such that the following holds:
\begin{equation}
	\begin{aligned}
		&t_i(w) \cdot \delta\phi_{ij}^{*}(w) + \alpha_i^{*}(w) - \beta_{ij}^{*}(w) = 0 \\
		&\sum_{j \in \mathcal{O}(i)}\phi_{ij}^{*}(w) - 1 = 0 \\
		&\alpha_i^{*}(w) \ge 0\\ 
		&\alpha_i^{*}(\sum_{j \in \mathcal{O}(i)}\phi_{ij}^{*}(w) - 1) = 0 \\
		&\beta_{ij}^{*}(w)\phi_{ij}^{*}(w) = 0\\
		&\alpha_i^{*}(w) \ge 0 \mathrm{\text{ and }} \beta_{ij}^{*}(w) \ge 0
	\end{aligned}
\end{equation}
Therefore, we can obtain that for all $w \in \mathcal{W}$ and $i \notin D(w)$ with $t_i(w) > 0$:
\begin{equation}
	t_i(w) \cdot \delta\phi_{ij}^{*}(w)=
	\begin{cases}
		- \alpha_i^{*}(w) + \beta_{ij}^{*}(w), & \mathrm{\text{ if }} \phi_{ij}^{*}(w) = 0,\\
		- \alpha_i^{*}(w), & \mathrm{\text{ if }} \phi_{ij}^{*}(w) > 0.
	\end{cases} \label{w}
\end{equation}
We should note that \eqref{w} is exactly the same as the Theorem 1 of \cite{xi2008node}, however, we here can give the exact meaning of the value $t_i(w) \cdot \delta\phi_{ij}^{*}(w)$ which equals to the negative optimal multiplier $- \alpha_i^{*}(w)$ of the constraint $\sum_{j \in \mathcal{O}(i)} \phi_{ij}(w) = 1$, while \cite{xi2008node} only states that $t_i(w) \cdot \delta\phi_{ij}^{*}(w)$ equals to a constant. Furthermore, following the same proof procedure in Theorem 1, we can get that for all $w \in \mathcal{W}$ and $i \notin D(w)$ with $t_i(w) > 0$:
\begin{equation}
	t_i(w) \cdot \delta\phi_{ij}^{*}(w) = -\alpha_i^{*}(w), \quad \forall j \in \mathcal{O}(i).
\end{equation}
That is, the partial derivatives of each element $\phi_{ij}^{*}(w)$ of the  sectional optimal point $\bm{\phi}_i^{*}(w)$ are the same, and it has no incentive to change the decision at that point. Furthermore, the node $i$ only route the workload of session $w$ to the link $(i,j)$ that has the minimum subgradient. Such optimal condition is appropriate for providing a distributed solution where at each node $i$, it can update the routing variables $\bm{\phi}_i(w)$ for session $w$ only based on its own information $\delta\phi_{ij}(w),  \forall j \in \mathcal{O}(i)$.

\subsection{The proof of Theorem 4} \label{appendix D}
We should point out that the proof of Theorem 4 is quite similar to that of Theorem 2. However, in Theorem 2 we apply mirror ascent to maximize the concave function $U(\Lambda, \bm{\phi}^{*}(\Lambda))$ while in Theorem 4 we apply mirror descent to minimize the convex function $D(\Lambda^{t},\bm{\phi})$. Recall the subgradient mirror descent rule
\begin{equation}
	\bm{\phi}^{k+1} = \mathop{\arg\min}\limits_{\bm{\phi} \in \mathcal{H}(\bm{\phi})} \Big\{ D(\bm{\phi}^k) + \langle \bm{t\cdot\delta\phi}^k, \bm{\phi} - \bm{\phi}^k\rangle  + \frac{1}{\eta_k}\Delta_{\psi}(\bm{\phi}, \bm{\phi}^k)
	\Big\}, \label{sm}
\end{equation}
where $\bm{t\cdot\delta\phi}^k$ is the subgradient (marginal cost). Similarly, the update rule \eqref{sm} can be interpreted as follows. First approximate $D$ around $\bm{\phi}^k$ by the first-order Taylor expansion 
\begin{equation}
	D(\bm{\phi}) \approx D(\bm{\phi}^k) + \langle \bm{t\cdot\delta\phi}^k, \bm{\phi} - \bm{\phi}^k\rangle.
\end{equation}
Then penalize the displacement by $\frac{1}{\eta_k}\Delta_{\psi}(\bm{\phi}, \bm{\phi}^k)$. The first order optimality condition for \eqref{sm} is 
\begin{align}
	&\bm{t\cdot\delta\phi}^k + \frac{1}{\eta_k}(\nabla\psi(\bm{\phi}^{k+1}) - \nabla\psi(\bm{\phi}^k)) = 0 \label{sn}\\
	\Longleftrightarrow \quad & \nabla\psi(\bm{\phi}^{k+1} = \nabla\psi(\bm{\phi}^k) - \eta_k \bm{t\cdot\delta\phi}^k  \notag\\
	\Longleftrightarrow \quad & \bm{\phi}^{k+1} = (\nabla\psi)^{-1}(\nabla\psi(\bm{\phi}^k) - \eta_k \bm{t\cdot\delta\phi}^k).
\end{align}
Therefore, the update rule of $\bm{\phi}^{k+1}$ has the following closed form 
\begin{equation*}
	\phi_{ij}^{k+1}(w) = \frac{\phi_{ij}^{k}(w)\exp(-\eta_k \delta\phi_{ij}^k(w))}{\sum_{j}\phi_{ij}^{k}(w)\exp(-\eta_k \delta\phi_{ij}^k(w))} 
\end{equation*}
which is exactly the Line 8 of OMA-RT algorithm.

We now show that after every iteration, the total network cost would decrease, that is 
\begin{equation}
	D(\bm{\phi}^{k+1}) \le D(\bm{\phi}^k), \quad \forall k \in \{1,\ldots,K\}.
\end{equation}
\begin{figure*}[t]
	\centering
	\begin{minipage}[t]{0.24\linewidth}
		\centering
		\includegraphics[width=1.75in]{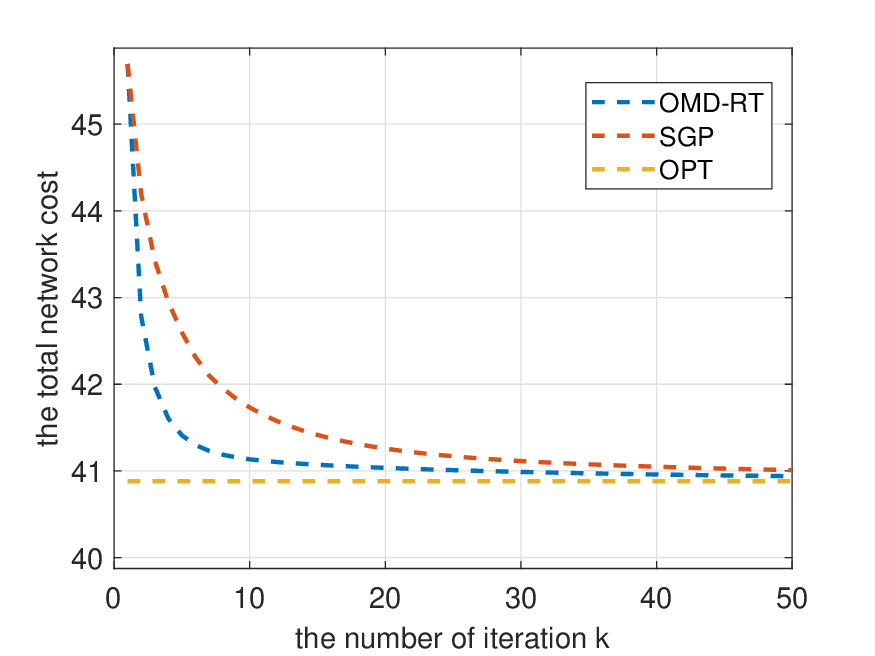}
		\caption{Abilene Topology.}
	\end{minipage}%
	\begin{minipage}[t]{0.24\linewidth}
		\centering
		\includegraphics[width=1.75in]{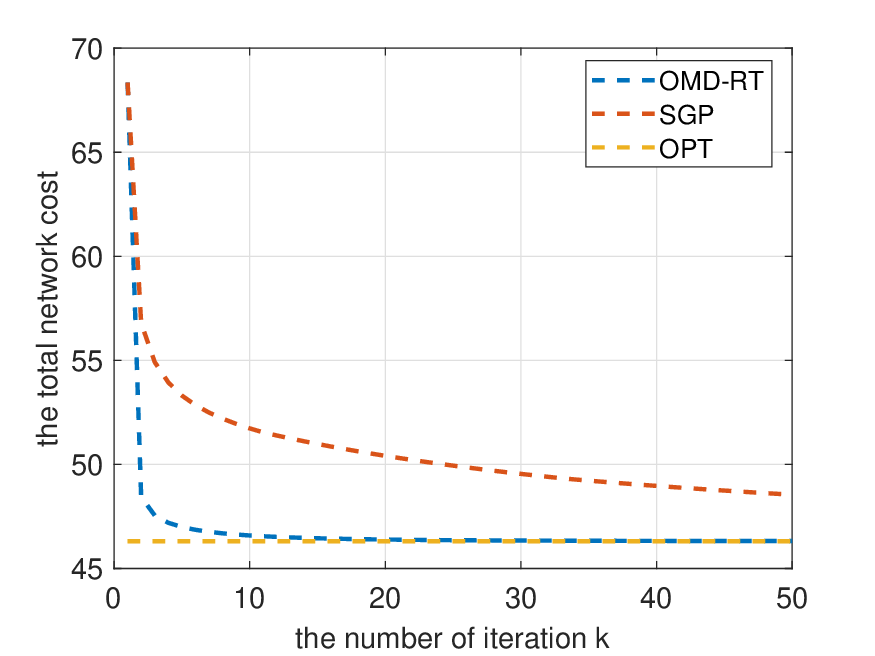}
		\caption{Tree Topology.}
	\end{minipage}%
	\begin{minipage}[t]{0.24\linewidth}
		\centering
		\includegraphics[width=1.75in]{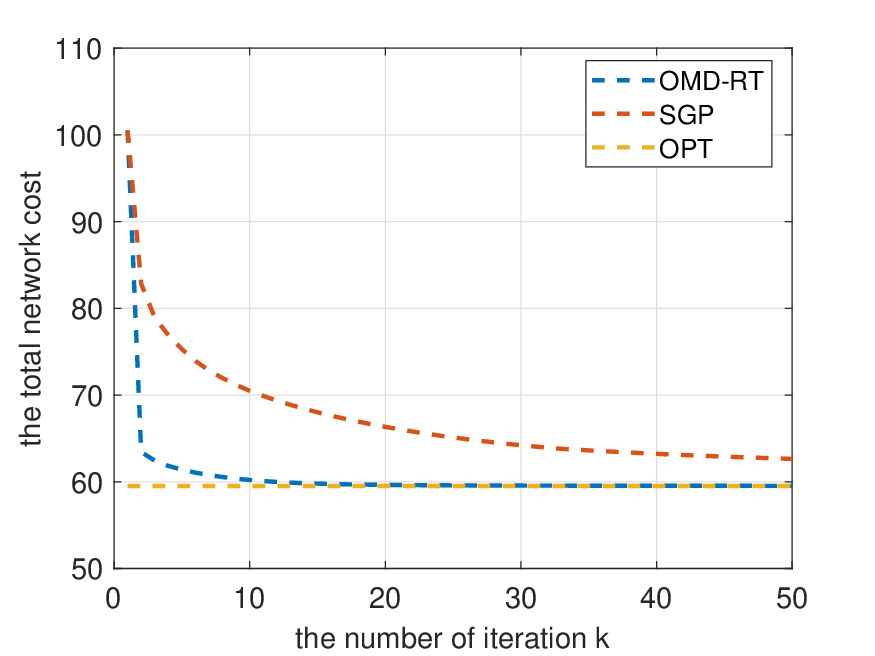}
		\caption{Fog Topology.}
	\end{minipage}
	\begin{minipage}[t]{0.24\linewidth}
		\centering
		\includegraphics[width=1.75in]{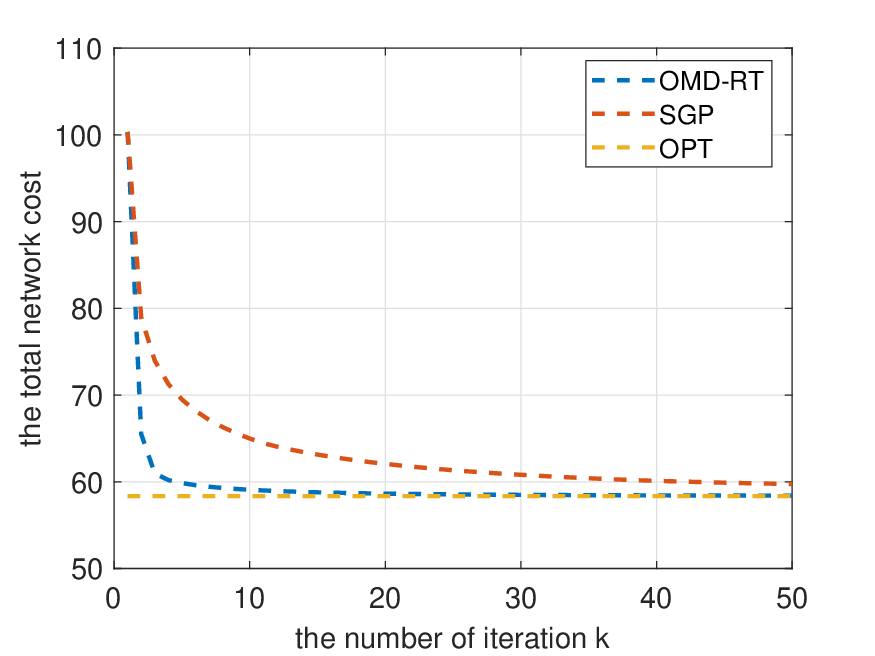}
		\caption{GEANT Topology.}
	\end{minipage}
\end{figure*}
Recall that $D$ has $L_D$-Lipschitz continuous gradient:
\begin{equation}
	D(\bm{\phi}) \le D(\bm{\phi}^k) + \langle \bm{t\cdot\delta\phi}^k, \bm{\phi} - \bm{\phi}^k \rangle + \frac{L_D}{2}||\bm{\phi} - \bm{\phi}^k||^2.
\end{equation}
Since $\psi(\bm{\phi})$ is $c$-strongly convex, we have 
\begin{equation}
	\Delta_{\psi}(\bm{\phi} - \bm{\phi}^k) \ge \frac{c}{2}||\bm{\phi} - \bm{\phi}^k||^2.
\end{equation}
Therefore, when we set $\eta_k \le \frac{c}{L_D}$, we can get
\begin{equation}
	\begin{aligned}
	 D(\bm{\phi}) &\le D(\bm{\phi}^k) + \langle \bm{t\cdot\delta\phi}^k, \bm{\phi} - \bm{\phi}^k \rangle + \frac{L_D}{2}||\bm{\phi} - \bm{\phi}^k||^2 \\
		&\le D(\bm{\phi}^k) + \langle \bm{t\cdot\delta\phi}^k, \bm{\phi} - \bm{\phi}^k \rangle + \frac{c}{2\eta_t}||\bm{\phi} - \bm{\phi}^k||^2 \\
		&\le D(\bm{\phi}^k) + \langle \bm{t\cdot\delta\phi}^k, \bm{\phi} - \bm{\phi}^k \rangle + \frac{1}{\eta_t}\Delta_{\psi}(\bm{\phi} - \bm{\phi}^k)
	\end{aligned}
\end{equation}
Finally, since $\bm{\phi}^{k+1}$ is the minimizer of the right-hand-side of above inequation, we have
\begin{equation}
	\begin{aligned}
		&D(\bm{\phi}^{k+1}) \le D(\bm{\phi}^k) + \langle \bm{t\cdot\delta\phi}^k, \bm{\phi}^{k+1} - \bm{\phi}^k \rangle  \\
		&\qquad \qquad \qquad \qquad \qquad + \frac{1}{\eta_k}\Delta_{\psi}(\bm{\phi}^{k+1} - \bm{\phi}^k)\\
		\Longleftrightarrow & D(\bm{\phi}^{k+1}) - D(\bm{\phi}^k) \le \langle \bm{t\cdot\delta\phi}^k, \bm{\phi}^{k+1} - \bm{\phi}^k \rangle  \\
		&\qquad \qquad \qquad \qquad \qquad \quad + \frac{1}{\eta_k}\Delta_{\psi}(\bm{\phi}^{k+1} - \bm{\phi}^k)\\
		\Longleftrightarrow &  D(\bm{\phi}^{k+1}) - D(\bm{\phi}^k) \le 0.
	\end{aligned}	
\end{equation}
The last inequation is because the function $\langle \bm{t\cdot\delta\phi}^k, \bm{\phi} - \bm{\phi}^k \rangle + \frac{1}{\eta_k}\Delta_{\psi}(\bm{\phi} - \bm{\phi}^k)$ has a value 0 at point $\bm{\phi}=\bm{\phi}^k$, and $\bm{\phi}^{k+1}$ is a minimizer of it, then we have $\langle \bm{t\cdot\delta\phi}^k, \bm{\phi}^{k+1} - \bm{\phi}^k \rangle + \frac{1}{\eta_k}\Delta_{\psi}(\bm{\phi}^{k+1} - \bm{\phi}^k) \le 0$.

We continue to analyze the rate of convergence of subgradient mirror descent in Algorithm 2. It takes four steps.\\
1. Bounding on a single update
\begin{equation}
	\begin{aligned}
		\Delta_{\psi}(\bm{\phi}^{*}, \bm{\phi}^{k+1}) &\le \Delta_{\psi}(\bm{\phi}^{*}, \bm{\phi}^k) + \eta_k(D(\bm{\phi}^{*}) - D(\bm{\phi}^{k+1}) \\& + \frac{L_D}{2}||\bm{\phi}^k-\bm{\phi}^{k+1}||^2) 
		- \frac{c}{2}||\bm{\phi}^k-\bm{\phi}^{k+1}||^2 \label{sp}
	\end{aligned}	
\end{equation}
By setting $\eta_k=\frac{c}{L_D}$, we can get 
\begin{equation}
	\Delta_{\psi}(\bm{\phi}^{*}, \bm{\phi}^{k+1}) \le \Delta_{\psi}(\bm{\phi}^{*}, \bm{\phi}^k) - \frac{c}{L_D}(D(\bm{\phi}^{k+1}) -D(\bm{\phi}^{*}) )
\end{equation}\\
2. Telescope over $k = 1, \ldots, K$:
\begin{equation}
	\Delta_{\psi}(\bm{\phi}^{*}, \bm{\phi}^{k+1}) \le \Delta_{\psi}(\bm{\phi}^{*}, \bm{\phi}^1) - \sum_{k=1}^{K}\frac{c}{L_D}(D(\bm{\phi}^{k+1})-D(\bm{\phi}^{*})).
\end{equation}
3. Bounding the $\Delta_{\psi}(\bm{\phi}^{*}, \bm{\phi}^1)$ with $R_D^2$ and rearrange: 
\begin{equation}
	\sum_{k=2}^{K+1}\frac{c}{L_D}(D(\bm{\phi}^k)-D(\bm{\phi}^{*})) \le R_D^{2} 
\end{equation}
By definition, $R_D^2 \triangleq \max_{\bm{\phi} \in \mathcal{H}(\bm{\phi})} \Delta_{\psi}(\bm{\phi}, \bm{\phi}^1)$ is the diameter of the decision space $\mathcal{H}(\bm{\phi})$ with the measure function $\psi$.\\
4. Denote $\epsilon_k \triangleq  D(\bm{\phi}^k)-D(\bm{\phi}^{*})$ and rearrange
\begin{equation}
	\min_{k \in \{2,\ldots,K+1\}} \epsilon_k \le \frac{L_DR_D^2}{cK}.
\end{equation}

\subsection{The proof of Theorem 5} \label{appendix E}
\cite{doi:10.1137/19M127375X} has proved that the single-loop gradient ascent descent algorithm can achieve the rate $O(\frac{1}{t})$ for a max-min problem in a smooth concave-convex setting if both the gradient ascent and gradient descent iteration guarantee the convergence rate $O(\frac{1}{t})$. Although in \cite{doi:10.1137/19M127375X}, the authors adopted the proximal point method to do the gradient ascent and descent and the proof of the convergence rate also based on the proximal point method, we are convinced of our proposed mirror method based gradient ascent descent also exhibits the $O(\frac{1}{t})$ convergence rate. 

For the preciseness of the proof, here we provide another way to prove the $O(\frac{1}{t})$ convergence rate using the Lyapunov functions. Consider the following two Lyapunov functions
\begin{equation}
	\begin{aligned}
		V_1(\Lambda) &= \max_{\Lambda} \min_{\bm{\phi}} U(\Lambda,\bm{\phi}(\Lambda)) - \min_{\bm{\phi}} U(\Lambda,\bm{\phi}(\Lambda)),\\
		V_2(\Lambda,\bm{\phi}) &= U(\Lambda,\bm{\phi}(\Lambda)) - \min_{\bm{\phi}} U(\Lambda,\bm{\phi}(\Lambda)).
	\end{aligned}
\end{equation}
where it is obvious to see that $V_1$ and $V_2$ are nonnegative. Define 
\begin{equation}
	V(\Lambda,\bm{\phi}) = V_1(\Lambda) + V_2(\Lambda,\bm{\phi}).
\end{equation}
The value of $V(\Lambda,\bm{\phi})$ is decreasing when using the gradient ascent descent algorithm, and finally converges to zero at the optimal point $(\Lambda^{*},\bm{\phi}^{*}(\Lambda^{*}))$. According to Theorem 2 and 4, we know that both $V_1(\Lambda)$ and $V_2(\Lambda,\bm{\phi})$ exhibit the $O(\frac{1}{t})$ convergence rate, and therefore the value of $V(\Lambda,\bm{\phi})$ also shares the $O(\frac{1}{t})$ convergence rate.
\subsection{Additional simulation results} \label{appendix F}
Table 2 shows the detailed parameters setup of different network topology where $|\mathcal{N}|$ is the number of nodes, $|\mathcal{E}|$ is the number of links, and $\bar{C}_{ij}$ is the mean link capacity. The convergence rate performance of OMD-RT is shown in Figure 11 to 14.
\begin{table}[h!]
	\begin{center}
		\caption{Simulated Network Scenarios}
		\begin{tabular}{|c|ccc|} 
			\hline
			Network &  & Parameters  & \\
			Topology & $|\mathcal{N}|$ & $|\mathcal{E}|$ & $\bar{C}_{ij}$\\
			\hline 
			Abilene & 11 & 14 & 15 \\
			Balanced-tree & 14 & 23 & 10 \\
			Fog & 15 & 30 & 10 \\
			GEANT & 22 & 33 & 10 \\
			\hline
		\end{tabular}
	\end{center}
\end{table}

%

\bibliographystyle{IEEEtran}
\bibliography{ref}

 




\vfill

\end{document}